\documentstyle[prb,eqsecnum,aps]{revtex}
\input epsf
\begin{document}
\draft
\title{Scattering of elastic waves by periodic arrays
of spherical bodies}
\author{I.~E.~Psarobas\cite{ntua}
 and N.~Stefanou}
\address{Section of Solid State Physics, University of Athens,
Panepistimioupolis,\\ GR-157 84, Athens, Greece}
\author{A.~Modinos}
\address{Department of Physics, National Technical University of Athens,
Zografou Campus,\\ GR-157 73, Athens, Greece}
\date{\today}
\maketitle
\begin{abstract}
{\bf We develop a formalism for the calculation of the frequency band
structure of a phononic crystal consisting of non-overlapping elastic
spheres, characterized by Lam\'e coefficients which may be complex
and frequency dependent, arranged periodically in a host medium with
different mass density and Lam\'e coefficients. We view the crystal
as a sequence of planes of spheres, parallel to and having the two
dimensional periodicity of a given crystallographic plane, and obtain
the complex band structure of the infinite crystal associated with
this plane. The method allows one to calculate, also, the
transmission, reflection, and absorption coefficients for an elastic
wave (longitudinal or transverse) incident, at any angle, on a slab
of the crystal of finite thickness. We demonstrate the efficiency of
the method by applying it to a specific example.}
\end{abstract}
\section{Introduction}
\label{intro} The elastic properties of a locally homogeneous and
isotropic composite material are characterized by a mass density
$\rho$ and Lam\'e coefficients $\lambda$ and $\mu$ which vary in
space.~\cite{Landau} The composite materials, we shall be concerned
with in this paper consist of homogeneous particles (solid or fluid
inclusions the dimensions of which must be large enough in order for
a macroscopic description of their elastic properties to be valid)
distributed periodically in a host medium characterized by different
mass density and Lam\'e coefficients. We assume, throughout this
paper, that the particles do not overlap with each other (cermet
topology~\cite{Ashcroft}). The alternative case, when the particles
connect with each other to form a continuous network is also
interesting but will not concern us here. When identical particles
are distributed periodically in a host medium, the composite material
may be referred to as a phononic crystal. In this case the mass
density and the Lam\'e coefficients vary periodically in space:
\begin{equation}
\rho({\bf r}+{\bf R}_n)=\rho({\bf r}), \;\; \mu({\bf r}+{\bf
R}_n)=\mu({\bf r}), \;\; \lambda({\bf r}+{\bf R}_n)=\lambda({\bf r}),
\label{eq:periodic}
\end{equation}
where $\{{\bf R}_n\}$ denotes a Bravais lattice.

In recent years there has been a growing interest in the study of
phononic crystals which is inspired to a large degree by
corresponding work in photonic crystals.~\cite{Joannop,Soukoulis}
These are composite materials with a dielectric function which varies
periodically in space. A typical example: identical particles, large
enough to be describable by a macroscopic dielectric function, are
arranged periodically in a host material with a different dielectric
function. Photonic crystals have many interesting properties both in
relation to basic physics and technological applications. In
particular, the existence of absolute frequency gaps (photonic gaps)
in certain such crystals, i.e., regions of frequency over which
electromagnetic (EM) waves can not exist within the crystal, has
attracted a lot of attention, mainly because of promising
applications in optoelectronics, as pointed out initially by
Yablonovich.~\cite{Yabl1} In principle, one can design a perfect
mirror, non-absorbing over a selected region of frequency
(corresponding to a photonic gap), a non-absorbing resonance cavity,
etc..~\cite{Yabl2} A number of theoretical calculations, predict the
existence of such gaps for appropriately designed photonic crystals,
but so far only crystals which exhibit gaps up to the infrared region
have been constructed.~\cite{PT99} However, progress to higher
frequencies is expected in the near future. In relation to basic
physics, photonic crystals are interesting in a number of
ways.\cite{SJ91} For example, they can be the starting point in a
process of gradual introduction of disorder and a study of consequent
phenomena, including Anderson localization.~\cite{PSheng}

Now, phononic crystals have properties which mirror those of photonic
crystals and corresponding applications too
.~\cite{Sig2,Kushw1,Kushw2,Sprik98,Kushw3,Kushw4,Sig3,Vasseur,Sanchez,Torres}
With an appropriate choice of the parameters involved one may obtain
phononic crystals with absolute frequency gaps (phononic gaps) in
selected regions of frequency. An elastic wave, whose frequency lies
within an absolute gap of a phononic crystal, will be completely
reflected by it; from which follows the possibility of constructing
non-absorbing mirrors of elastic waves and vibration-free cavities
which might be very useful in high-precision mechanical systems
operating in a given frequency range. And in relation to basic
physics, one can use elastic waves to study phenomena such as those
associated with disorder,~\cite{Localiz} in more or less the same
manner as with EM waves.

There are, however, some essential differences between EM and elastic
waves and this means that the normal modes of the elastic field in a
phononic crystal are in some ways quite different from those of the
EM field in a photonic crystal. In a homogeneous isotropic medium the
elastic waves can, in general, be purely longitudinal (in which case
the displacement vector ${\bf u}({\bf r})$ satisfies the condition
$\nabla \times {\bf u}=0$) or purely transverse (in which case
$\nabla \cdot {\bf u}=0$). In a phononic crystal this is no longer
the case and a normal mode usually has a longitudinal and a
transverse component. One expects that because of this coupling
between longitudinal and transverse waves, it will be more difficult
to obtain absolute frequency gaps in a phononic crystal. We recall
that the normal modes of the EM field in a photonic crystal are
exclusively transverse. On the other hand there are, in general, more
parameters relevant to the determination of phononic gaps than there
are in the determination of photonic gaps. In  the case of a binary
system (consisting of material 2 distributed in material 1) we have
for photonic crystals two independent parameters: the ratio of the
dielectric functions of the two materials $\epsilon_2/\epsilon_1$;
and the fractional volume occupied by material 2, to be denoted by
$f$. For phononic crystals there are five independent parameters:
$\mu_2/\mu_1$, $\lambda_2/\lambda_1$, $\rho_2/\rho_1$,
$\mu_2/\lambda_1$ and $f$; where $\rho_j$, $\lambda_j$, $\mu_j$
denote the mass density, and the Lam\'e coefficients of material
$j=1,2$.

We have, so far, implicitly assumed that the Lam\'e coefficients
describing the constituent materials of the phononic crystals are all
different from zero, real quantities, and constant (independent of
the frequency). But this is not always the case. The phononic crystal
may consist, for example, of solid particles (material 2) arranged
periodically (at least approximately) in a liquid (material 1). If
the liquid is a normal fluid, like water, $\mu_1=0$ and the
transverse sound in the liquid is suppressed. This, however, is not
the case for a viscous fluid. The role of shear viscosity in phononic
crystals has been pointed out by Sprik and Wegdam. ~\cite{Sprik98}
Shear viscosity is equally important in phononic crystals consisting
of liquid particles in a solid host background (liquid-containing
porous
 solids~\cite{Biot,Schwartz,Marqusee,Johnson}).
Colloidal suspensions of solid spheres in a liquid, also, have
interesting acoustic properties.~\cite{Colloid} Finally, it may be of
some interest to consider composite materials consisting of two
liquids (e.g. drops of oil in water) although in this case a periodic
arrangement of the drops can only be a rough approximation to the
real system. It appears that acoustic gaps are easily obtained in
three-dimensional (3D) fluid-fluid composites, when
$\mu_1=\mu_2=0$.~\cite{Kushw1}

The few calculations published so far relating to 3D phononic
crystals deal, almost exclusively, with the frequency band structure
of these crystals, which is obtained via a plane-wave expansion of
the displacement field.~\cite{Sig2,Kushw1,Kushw2,Sprik98} On the
other hand, a lot of theoretical and experimental work has been done
on systems with two-dimensional (2D) periodicity, with translational
invariance along the third dimension. A typical example of such
systems consists of a set of long identical cylinders parallel to the
$z$ direction, crossing the $xy$ plane at the sites of a 2D lattice.
By considering waves propagating normal to the cylinders, the problem
is reduced to two
dimensions.~\cite{Kushw3,Kushw4,Sig3,Vasseur,Sanchez,Torres} The
above investigations have shown that phononic gaps are possible in
both 2D and 3D systems.

Although knowing the frequency band structure of a phononic crystal
is very useful, more is required for a full interpretation and
analysis of the experimental data. In an experiment one usually
measures the reflection and/or transmission coefficients of an
acoustic/elastic wave incident on a slab of the phononic crystal, and
consequently theory should be able to provide reliable estimates of
these, the experimentally measured quantities, as well. The so-called
on-shell methods developed in relation to photonic crystals can do
exactly that, besides an accurate evaluation of the frequency band
structure.~\cite{Pendry92,SKM92,Ohtaka} In these methods one
determines for a given frequency $\omega$ and a given reduced
wavevector, ${\bf k}_{\|}$, parallel to a given crystallographic
plane of the crystal, the Bloch-wave solutions of the elastic field
of the infinite crystal; these consist of propagating and evanescent
waves. The propagating waves constitute the normal modes of the
infinite phononic crystal. The evanescent waves do not represent real
waves, they are mathematical entities which enter directly or
indirectly (depending on the method of calculation) into the
evaluation of the reflection and transmission coefficients of a wave,
with given $\omega$ and ${\bf k}_{\|}$, incident on a slab of the
crystal parallel to the given crystallographic plane. On-shell
methods have certain advantages over the plane-wave method, even if
one is only interested in the frequency band structure and the
corresponding normal modes of vibration of the infinite phononic
crystal. In an on-shell method one can easily allow the Lam\'e
coefficients of any of the constituent materials of the crystal to
depend on the frequency, as is necessary in some cases, without any
difficulty, which is not the case with the plane-wave method. And, as
a rule, on-shell methods are computationally more
efficient.~\cite{Pendry92,Leung2}

The on-shell method we describe in the present paper is analogous to
that which some of us have developed for photonic
crystals.~\cite{SKM92} It applies to systems which consist of
non-overlapping spherical particles arranged periodically in a host
medium characterized by different mass density and Lam\'e
coefficients. Sections \ref{Multexp} to \ref{band} are devoted to the
development of the formalism.~\cite{footnote1} In section
\ref{applic} we demonstrate the applicability of the method on a
specific system: an fcc crystal of silica spheres in ice. Finally the
last section concludes this article.
\section{Multipole expansion of the elastic field}
\label{Multexp} The displacement vector ${\bf U}({\bf r},t)$, in a
homogeneous elastic medium of mass density $\rho$ and Lam\'e
coefficients $\lambda$, $\mu$, satisfies the equation~\cite{Landau}
\begin{equation}
\left(\lambda+2\mu\right)\nabla\left(\nabla\cdot {\bf U}\right)-\mu
\nabla\times\left(\nabla\times {\bf U} \right) -\rho
{\partial}^2_t{\bf U}=0 . \label{eq:tdep}
\end{equation}
In the case of a harmonic elastic wave of angular frequency $\omega$,
we have
\begin{equation}
{\bf U}({\bf r},t)={\rm Re}\left[{\bf u}({\bf r}){\rm exp}(-{\rm
i}\omega t)\right] , \label{eq:harmonic}
\end{equation}
and Eq.~(\ref{eq:tdep}) reduces to the following time-independent
form
\begin{equation}
\left(\lambda+2\mu\right)\nabla\left(\nabla\cdot {\bf u}\right)-\mu
\nabla\times\left(\nabla\times {\bf u} \right)+\rho \omega^2{\bf u}=0
. \label{eq:tindep}
\end{equation}
We note that for ordinary elastic media the Lam\'e coefficients are
real numbers. Media where loss is possible, assuming the time
dependence given in Eq.~(\ref{eq:harmonic}), are described by complex
Lam\'e coefficients~\cite{Brill}:
\begin{equation}
\lambda=\lambda_e-{\rm i} \omega \lambda_v ,\,\,\, \mu=\mu_e-{\rm i}
\omega \mu_v. \label{eq:visc}
\end{equation}

The most general solution of Eq.~(\ref{eq:tindep}) consists of two
elastic waves which propagate independently. These are: a
longitudinal (irrotational) wave, which satisfies the equations
\begin{equation}
\nabla^2{\bf u}+q_l^2{\bf u}=0 ,\;\nabla\times{\bf u}=0 ,
\label{eq:acoustic}
\end{equation}
where $q_l=\omega/c_l$, $c_l=\sqrt{(\lambda+2\mu)/\rho}$ being the
speed of propagation of this wave; and a transverse (divergenceless)
wave, which satisfies the equations
\begin{equation}
\nabla^2{\bf u}+q_t^2{\bf u}=0,\;\; \nabla\cdot{\bf u}=0 ,
\label{eq:em}
\end{equation}
where $q_t=\omega/c_t$, $c_t=\sqrt{\mu/\rho}$ being the speed of
propagation of this wave.

In the present paper we shall use, besides the more familiar
solutions of Eqs.~(\ref{eq:acoustic}) and (\ref{eq:em}) representing
longitudinal and transverse plane elastic waves (see
Eq.~(\ref{eq:upw}) below), the so-called spherical-wave solutions of
these equations. A complete set of spherical-wave solutions of
Eq.~(\ref{eq:acoustic}), known as irrotational vector wave functions,
is given by~\cite{Chew}
\begin{mathletters}
\begin{equation}
{\bf u}^L_{\ell m}({\bf r})=\frac{1}{q_l} \nabla
\left[f_{\ell}(q_lr)Y_{\ell}^m(\hat{{\bf r}})\right]\,, \label{eq:uL}
\end{equation}
where $f_{\ell}$ may be any linear combination of the spherical
Bessel function, $j_{\ell}$, and the spherical Hankel function,
$h_{\ell}^+$. $Y_{\ell}^m(\hat{{\bf r}})$ are the usual spherical
harmonics, with $\hat{{\bf r}}$ denoting the angular variables
$(\theta ,\phi)$ of ${\bf r}$ in a system of spherical coordinates.

A complete set of spherical-wave solutions of Eq.~(\ref{eq:em}) is
given by~\cite{Chew}
\begin{equation}
{\bf u}^M_{\ell m}({\bf r})=f_{\ell}(q_tr){\bf X} _{\ell m}(\hat{{\bf
r}}) \label{eq:uM}
\end{equation}
and
\begin{equation} {\bf u}^N_{\ell m}({\bf r})=\frac{{\rm
i}}{q_t}\nabla\times f_{\ell}(q_tr){\bf X} _{\ell m}(\hat{{\bf
r}})\,, \label{eq:uN}
\end{equation}
\end{mathletters}
which are also known as solenoidal vector wave functions. The vector
spherical harmonics, denoted by ${\bf X} _{\ell m}(\hat{{\bf r}})$,
are defined by
\begin{mathletters}
\begin{equation}
\sqrt{\ell (\ell+1)}{\bf X}_{\ell m}(\hat{{\bf r}})={\bf
L}Y_{\ell}^m(\hat{{\bf r}})\equiv -{\rm i}{\bf r}\times\nabla
Y_{\ell}^m(\hat{{\bf r}})\;. \label{eq:Xlm}
\end{equation}
By definition ${\bf X}_{00}(\hat{{\bf r}})=0$; for $\ell\geq1$ we
have
\begin{eqnarray}
\sqrt{\ell(\ell+1)}{\bf X}_{\ell m}(\hat{{\bf r}})= &&
\left[\alpha_{\ell}^{-m}\cos \theta \;e^{{\rm
i}\phi}\;Y_{\ell}^{m-1}(\hat{{\bf r}})-m\sin \theta\;
Y_{\ell}^{m}(\hat{{\bf r}}) +\alpha_{\ell}^m \cos \theta \;e^{-{\rm
i}\phi}\;Y_{\ell}^{m+1}(\hat{{\bf r}})\right]\hat{{\bf e}}_{\theta}
\nonumber \\ && +\;{\rm i}\left[\alpha_{\ell}^{-m}\;e^{{\rm
i}\phi}\;Y_{\ell}^{m-1}(\hat{{\bf r}})-\alpha_{\ell}^m\;e^{-{\rm
i}\phi}\;Y_{\ell}^{m+1}(\hat{{\bf r}})\right]\hat{{\bf e}}_{\phi}\;,
\label{eq:CartXlm}
\end{eqnarray}
where
\begin{eqnarray}
\alpha_{\ell}^m=\frac{1}{2}[(\ell-m)(\ell+m+1)]^{1/2}\;,
\label{eq:alm}
\end{eqnarray}
\end{mathletters}
and $\hat{{\bf e}}_{\theta}$, $\hat{{\bf e}}_{\phi}$, are the usual
polar and azimuthal unit vectors, respectively, in the chosen system
of spherical coordinates.

The most general displacement field can be written as a linear sum of
the spherical waves given by Eqs.~(\ref{eq:uL})-(\ref{eq:uN}), as
follows
\begin{eqnarray}
{\bf u}({\bf r})=\sum_{\ell m} \left\{{\rm a}_{\ell m}^M
f_{\ell}(q_tr){\bf X} _{\ell m}(\hat{{\bf r}})+{\rm a}_{\ell
m}^N\frac{{\rm i}}{q_t}\nabla\times f_{\ell}(q_tr){\bf X} _{\ell
m}(\hat{{\bf r}})+{\rm a}_{\ell m}^L \frac{1}{q_l} \nabla
\left[f_{\ell}(q_lr)Y_{\ell}^m(\hat{{\bf r}})\right]\right\}\;,
\label{eq:multexp}
\end{eqnarray}
where ${\rm a}_{\ell m}^{P}$, $P=M,N,L$, are coefficients to be
determined.
\section{Scattering of a plane wave by a sphere}\label{sphscat}
A plane elastic wave, of wavevector ${\bf q}$, propagating in a
homogeneous elastic medium has the form
\begin{equation}
{\bf u}_{\rm in}({\bf r})={\bf u}_0({\bf q})\exp({\rm i}{\bf q}\cdot
{\bf r}), \label{eq:upw}
\end{equation}
with ${\bf u}_0({\bf q})=u_0({\bf q})\hat{{\bf e}}$, where $u_0$
denotes the magnitude and $\hat{{\bf e}}$, a unit vector, the
polarization of the displacement field. In the case of a longitudinal
plane wave we can write: ${\bf q}=q_l\hat{{\bf e}}_q$ and $\hat{{\bf
e}}=\hat{{\bf e}}_q$. Since the plane wave is finite everywhere, its
multipole expansion into spherical waves, according to
Eq.~(\ref{eq:uL}), involves only the radial functions
$j_{\ell}(q_lr)$; we have
\begin{eqnarray}
{\bf u}_{\rm in}({\bf r})=\sum_{\ell m} {\rm a}^{0L}_{\ell
m}\frac{1}{q_l} \nabla \left[j_{\ell}(q_lr)Y_{\ell}^m(\hat{{\bf
r}})\right] \,. \label{eq:ulinc}
\end{eqnarray}
One can easily show that the coefficients ${\rm a}^{0L}_{\ell m}$ are
given by
\begin{equation}
{\rm a}_{\ell m}^{0L}={\bf A}_{\ell m}^{0L}(\hat{{\bf q}}) \cdot {\bf
u}_0({\bf q})\,, \label{eq:a0L}
\end{equation}
where
\begin{equation} {\bf A}^{0L}_{\ell m}(\hat{{\bf q}})=4\pi {\rm
i}^{\ell+1}(-1)^{m+1}Y_{\ell}^{-m}(\hat{{\bf q}})\;\hat{{\bf e}}_q\,.
\label{eq:A0L}
\end{equation}
In the case of a transverse plane wave we have: ${\bf q}=q_t\hat{{\bf
e}}_q$ and $\hat{{\bf e}}\bot \hat{{\bf e}}_q$. Such a wave can be
written as a linear sum of the spherical waves given by
Eqs.~(\ref{eq:uM}) and (\ref{eq:uN}), and again involves only the
radial functions $j_{\ell}(q_tr)$; we have
\begin{eqnarray}
{\bf u}_{\rm in}({\bf r})=\sum_{\ell m} \biggl\{{\rm a}_{\ell m}^{0M}
j_{\ell}(q_tr){\bf X} _{\ell m}(\hat{{\bf r}})+{\rm a}_{\ell
m}^{0N}\frac{{\rm i}}{q_t}\nabla\times j_{\ell}(q_tr){\bf X} _{\ell
m}(\hat{{\bf r}})\biggr\}\;. \label{eq:utinc}
\end{eqnarray}
The coefficients ${\rm a}^{0P}_{\ell m}$, with $P=M,N$, can be
written as
\begin{equation}
{\rm a}^{0P}_{\ell m}={\bf A}_{\ell m}^{0P}(\hat{{\bf q}}) \cdot {\bf
u}_0({\bf q}) , \label{eq:a0P}
\end{equation}
where
\begin{eqnarray}
{\bf A}_{\ell m}^{0M}(\hat{{\bf q}}) && \;=\frac{4\pi {\rm
i}^{\ell}(-1)^{m+1}}{\sqrt{\ell (\ell +1)}}\nonumber \\ && \times
\biggl\{\left[\alpha_{\ell}^{m}\cos \theta \;e^{{\rm
i}\phi}\;Y_{\ell}^{-m-1}(\hat{{\bf q}})+m\sin \theta\;
Y_{\ell}^{-m}(\hat{{\bf q}}) +\alpha_{\ell}^{-m} \cos \theta
\;e^{-{\rm i}\phi}\;Y_{\ell}^{-m+1}(\hat{{\bf q}})\right]\hat{{\bf
e}}_{\theta} \nonumber
\\ && +{\rm i}\left[\alpha_{\ell}^{m}\;e^{{\rm i}\phi}\;Y_{\ell}^{-m-1}(\hat{{\bf
q}})-\alpha_{\ell}^{-m}\;e^{-{\rm i}\phi}\;Y_{\ell}^{-m+1}(\hat{{\bf
q}})\right]\hat{{\bf e}}_{\phi}\biggr\}\;, \label{eq:A0M}
\end{eqnarray}
and
\begin{eqnarray}
{\bf A}_{\ell m}^{0N}(\hat{{\bf q}})=&& \frac{4\pi {\rm
i}^{\ell}(-1)^{m+1}}{\sqrt{\ell (\ell +1)}}\biggl\{{\rm i}
\left[\alpha_{\ell}^{m}\;e^{{\rm i}\phi}\;Y_{\ell}^{-m-1}(\hat{{\bf
q}})-\alpha_{\ell}^{-m}\;e^{-{\rm i}\phi}\;Y_{\ell}^{-m+1}(\hat{{\bf
q}})\right]\hat{{\bf e}}_{\theta}\nonumber \\&&
-\left[\alpha_{\ell}^{m}\cos \theta \;e^{{\rm
i}\phi}\;Y_{\ell}^{-m-1}(\hat{{\bf q}})+m\sin \theta\;
Y_{\ell}^{-m}(\hat{{\bf q}}) +\alpha_{\ell}^{-m} \cos \theta
\;e^{-{\rm i}\phi}\;Y_{\ell}^{-m+1}(\hat{{\bf q}})\right]\hat{{\bf
e}}_{\phi}\biggr\}\;,\label{eq:A0N}
\end{eqnarray}
where $\theta$ and $\phi$ denote the angular variables of ${\bf q}$
in the chosen system of spherical coordinates.

We now consider a sphere of radius $S$, centered at the origin of
coordinates. We assume that the sphere, which has a uniform mass
density $\rho_s$, is embedded in a homogeneous medium of mass density
$\rho$; the wavenumbers of the elastic waves in the sphere
($q_{s\nu}$) and in the host medium ($q_{\nu}$), where $\nu=l$ or
$t$, are also different. When a plane wave is incident on the sphere,
it is scattered by it, so that the wavefield outside the sphere
consists of the incident wave and a scattered wave. Since the
scattered wave is outgoing at infinity, its expansion in spherical
waves is given by Eq.~(\ref{eq:multexp}) with $f_{\ell}=h^+_{\ell}$,
which has the asymptotic form appropriate to an outgoing spherical
wave: $h^+_{\ell}(qr)\approx (-{\rm i})^{\ell}\exp ({\rm i}qr)/{\rm
i}qr$ as $r\rightarrow\infty$. We have
\begin{eqnarray}
{\bf u}_{\rm sc}({\bf r})=\sum_{\ell m} \biggl\{{\rm a}_{\ell m}^{+M}
h^+_{\ell}(q_tr){\bf X} _{\ell m}(\hat{{\bf r}})+{\rm a}_{\ell
m}^{+N}\frac{{\rm i}}{q_t}\nabla\times h^+_{\ell}(q_tr){\bf X} _{\ell
m}(\hat{{\bf r}})+{\rm a}_{\ell m}^{+L} \frac{1}{q_l} \nabla
\left[h^+_{\ell}(q_lr)Y_{\ell}^m(\hat{{\bf r}})\right]\biggr\}\;.
\label{eq:scat}
\end{eqnarray}
The wavefield inside the sphere is given by Eq.~(\ref{eq:multexp})
with $f_{\ell}=j_{\ell}$, since it must be finite at the origin; we
have
\begin{eqnarray}
{\bf u}_{I}({\bf r})=\sum_{\ell m} \biggl\{{\rm a}_{\ell m}^{IM}
j_{\ell}(q_{st}r){\bf X} _{\ell m}(\hat{{\bf r}})+{\rm a}_{\ell
m}^{IN}\frac{{\rm i}}{q_{st}}\nabla\times j_{\ell}(q_{st}r){\bf X}
_{\ell m}(\hat{{\bf r}})+{\rm a}_{\ell m}^{IL} \frac{1}{q_{sl}}
\nabla \left[j_{\ell}(q_{sl}r)Y_{\ell}^m(\hat{{\bf
r}})\right]\biggr\}\;. \label{eq:inside}
\end{eqnarray}
The coefficients ${\rm a}_{\ell m}^{+P}$, ${\rm a}_{\ell m}^{IP}$,
$P=M,N,L$, in Eqs.~(\ref{eq:scat}) and (\ref{eq:inside}) are
determined by the requirement of continuity of the displacement
vector, ${\bf u}({\bf r})$, and of the surface traction,
$\mbox{\boldmath $\tau$}({\bf r})\equiv \overline{\mbox{\boldmath $
\sigma$}}({\bf r})\cdot \hat{{\bf r}}$, at the surface of the sphere;
$\overline{\mbox{\boldmath $ \sigma$}}({\bf r})$ denotes the stress
tensor. The components of the surface traction are given by (see,
e.g., Ref.~\onlinecite{Landau})
\begin{mathletters}
\begin{eqnarray}
\tau_r &=& \lambda \nabla \cdot {\bf u}+2\mu \partial_ru_r \\
\tau_{\theta} &=& \mu \left[\frac{1}{r}\partial_{\theta}u_r+
\partial_ru_{\theta}-\frac{u_{\theta}}{r}\right] \\
\tau_{\phi} &=& \mu \left[\frac{1}{r\sin \theta}\partial_{\phi}u_r+
\partial_ru_{\phi}-\frac{u_{\phi}}{r}\right].
\end{eqnarray}
\end{mathletters}
The continuity of $u_r$, $u_{\theta}$, $u_{\phi}$, $\tau_r$,
$\tau_{\theta}$, $\tau_{\phi}$ at the surface of the sphere allows us
to determine uniquely the coefficients ${\rm a}_{\ell m}^{+P}$
($P=M,N,L$) of the scattered wave, given by Eq.~(\ref{eq:scat}), and
the coefficients ${\rm a}_{\ell m}^{IP}$ of the wave inside the
sphere, given by Eq.~(\ref{eq:inside}), in terms of the known
coefficients ${\rm a}_{\ell m}^{0P}$ of the incident wave, given by
Eqs.~(\ref{eq:ulinc}) or (\ref{eq:utinc}). After some lengthy but
straightforward algebra one obtains (see, e.g.,
Ref.~\onlinecite{Brill})
\begin{equation}
{\rm a}_{\ell m}^{+P}=\sum_{P' \ell' m'} T_{\ell m;\ell' m'}^{PP'}\;
{\rm a}_{\ell' m'}^{0P'}\;. \label{eq:scat1}
\end{equation}
Explicit expressions for the nonzero elements of the {\bf T} matrix
in the case of a solid scatterer in a solid host are given in
Appendix~\ref{apxa}. Similar expressions for the cases involving a
liquid scatterer or host can be found in Ref.~\onlinecite{Brill}.
\section{Scattering by a plane of spheres}
\label{plane} We consider a plane of spheres at $z=0$: an array of
spheres centered on the sites of a 2D lattice specified by
\begin{equation}
{\bf R}_n=n_1{\bf a}_1+n_2{\bf a}_2\;, \label{eq:Rn}
\end{equation}
where ${\bf a}_1$ and ${\bf a}_2$ are primitive vectors in the $xy$
plane and $n_1,n_2=0,\pm1,\pm2,\pm3,\ldots$.

The corresponding 2D reciprocal lattice is obtained in the usual
manner~\cite{Pendry,Modinos} as follows
\begin{equation}
{\bf g}=m_1{\bf b}_1+m_2{\bf b}_2\;, \label{eq:g}
\end{equation}
where $m_1,m_2=0,\pm1,\pm2,\pm3,\ldots$ and ${\bf b}_1,{\bf b}_2$ are
defined by
\begin{equation}
{\bf b}_i\cdot {\bf a}_j=2\pi \delta_{ij}\;. \label{eq:badot}
\end{equation}

We now assume that a plane wave (it can be longitudinal or
transverse) is incident on the plane of spheres. We write the
displacement vector ${\bf u}_{{\rm in}}({\bf r})$ corresponding to it
as follows
\begin{equation}
{\bf u}_{\rm in}^{s'}({\bf r})=\sum_{i'} \left[u_{\rm
in}\right]^{s'}_{{\bf g}' i'} \exp ({\rm i}{\bf K}_{{\bf
g}'\nu'}^{s'} \cdot {\bf r}) \hat{{\bf e}}_{i'} \;, \label{eq:uinc}
\end{equation}
where $s'=+(-)$ corresponds to a wave incident on the plane of
spheres from the left (right); $\nu'$ specifies the polarization of
the incident wave: $q_{\nu'}=q_l=\omega/c_l$ for a longitudinal wave
and $q_{\nu'}=q_t=\omega/c_t$ for a transverse wave;
\begin{equation}
{\bf K}_{{\bf g}'\nu'}^{\pm}\equiv {\bf k}_{\|}+{\bf g}'\pm
\left[q_{\nu'}^2-({\bf k}_{\|}+{\bf g}')^2\right]^{1/2}\hat{{\bf
e}}_z\;, \label{eq:Kg}
\end{equation}
where $\hat{{\bf e}}_z$ is the unit vector along the $z$ axis, and we
have written the component of the incident wavevector parallel to the
plane of spheres as the sum of a reduced wavevector ${\bf k}_{\|}$,
which lies in the surface Brillouin zone (SBZ) of the given lattice,
and an appropriate reciprocal-lattice vector ${\bf g}'$. This is
always possible and it facilitates the subsequent calculation. For
$\nu'=l$, $i'=1$ denotes the only non-zero component of the
displacement vector, $\hat{{\bf e}}_1$ being the radial unit vector
along the direction of ${\bf K}_{{\bf g}'l}^{s'}$. For $\nu'=t$,
$i'=2,3$ denote the only non-zero components of the displacement
vector; $\hat{{\bf e}}_2,\hat{{\bf e}}_3$ being the polar and
azimuthal unit vectors, respectively, which are perpendicular to
${\bf K}_{{\bf g}'t}^{s'}$. In the same manner (as in
Eq.~(\ref{eq:Kg})) we define, for given ${\bf k}_{\|}$, $\omega$, a
wavevector ${\bf K}_{{\bf g}\nu}^s$ and the corresponding $\hat{{\bf
e}}_i$ for any ${\bf g}$ and any $\nu$. We remember that the $i=1$
component of the displacement vector is always associated with a
longitudinal plane wave ($\nu=l$), and that the $i=2,3$ components of
the displacement vector are always associated with a transverse plane
wave ($\nu=t$), so that the character (longitudinal or transverse) of
a given plane wave is automatically determined by the non-vanishing
components of the displacement vector associated with it, and need
not be stated explicitly in every case. When $({\bf k}_{\|}+{\bf
g})^2>q_{\nu}^2$ the corresponding wave decays to the right for
$s=+$, and to the left for $s=-$; and the corresponding unit vectors
$\hat{{\bf e}}_i$ become complex. Indeed, the unit vectors $\hat{{\bf
e}}_i$ are defined in a Cartesian system of coordinates as follows
\begin{mathletters}
\begin{equation}
\hat{{\bf e}}_1=\hat{{\bf e}}_x \sin \theta \cos \phi+\hat{{\bf e}}_y
\sin \theta \sin \phi+\hat{{\bf e}}_z \cos \theta \;, \label{eq:e1}
\end{equation}
where $\theta$ and $\phi$ denote the angular variables of ${\bf
K}_{{\bf g}l}^s$, and
\begin{eqnarray}
\hat{{\bf e}}_2 &=& \hat{{\bf e}}_x \cos \theta \cos \phi+\hat{{\bf
e}}_y \cos \theta \sin \phi-\hat{{\bf e}}_z \sin \theta \;,
\label{eq:e2}
\\
\hat{{\bf e}}_3 &=& -\hat{{\bf e}}_x \sin \phi+\hat{{\bf e}}_y \cos
\phi \;, \label{eq:e3}
\end{eqnarray}
\end{mathletters}
where $\theta$ and $\phi$ here denote the angular variables of ${\bf
K}_{{\bf g}t}^s$. We note that the $z$ component of ${\bf K}_{{\bf
g}\nu}^s$ (denoted by $K_{{\bf g}\nu z }^s$) is real if $({\bf
k}_{\|}+{\bf g})^2<q_{\nu}^2$ and imaginary if $({\bf k}_{\|}+{\bf
g})^2>q_{\nu}^2$. In the latter case, $\cos \theta_{{\bf K}_{{\bf
g}\nu}^s}$ in Eqs.~(\ref{eq:e1}) and (\ref{eq:e2}) is replaced by
$K_{{\bf g}\nu z}^s/q_{\nu}$ and $\sin \theta_{{\bf K}_{{\bf
g}\nu}^s}$ by $|{\bf k}_{\|}+{\bf g}|/q_{\nu}$, so that $\hat{{\bf
e}}_1$ and $\hat{{\bf e}}_2$ become complex.

Because of the 2D periodicity of the structure under consideration,
the wave scattered from it, when the wave given by
Eq.~(\ref{eq:uinc}) is incident upon it, has the form
\begin{eqnarray}
{\bf u}_{\rm sc}({\bf r}) &=& \sum_{\ell m} \biggl\{{\rm b}_{\ell
m}^{+M} \sum_{{\bf R}_n} \exp ({\rm i}{\bf k}_{\|}\cdot {\bf R}_n)
h^+_{\ell}(q_tr_n){\bf X} _{\ell m}(\hat{{\bf r}}_n)\nonumber
\\ & & +{\rm b}_{\ell m}^{+N}\frac{{\rm i}}{q_t}\nabla \times
 \sum_{{\bf R}_n}
\exp ({\rm i}{\bf k}_{\|}\cdot {\bf R}_n) h_{\ell}^+(q_tr_n){\bf X}
_{\ell m}(\hat{{\bf r}}_n)\nonumber \\ & & +{\rm b}_{\ell
m}^{+L}\frac{1}{q_l}\nabla \sum_{{\bf R}_n} \exp ({\rm i}{\bf
k}_{\|}\cdot {\bf R}_n) h^+_{\ell}(q_lr_n)Y_{\ell}^{m}(\hat{{\bf
r}}_n)\biggr\}\;, \label{eq:multscat}
\end{eqnarray}
where ${\bf r}_n={\bf r}-{\bf R}_n$. We note that $\exp[{\rm i}({\bf
k}_{\|}+{\bf g})\cdot{\bf R}_n]=\exp({\rm i}{\bf k}_{\|}\cdot{\bf
R}_n)$ because of Eq.~(\ref{eq:badot}). Eq.~(\ref{eq:multscat}) tells
us that the scattered wave is a sum of outgoing spherical waves
centered on the spheres of the plane, and that the wave scattered
from the sphere at ${\bf R}_n$ differs from that scattered from the
sphere at the origin (${\bf R}_n={\bf 0}$) only by the phase factor
$\exp ({\rm i}{\bf k}_{\|}\cdot {\bf R}_n)$. We note the presence in
the scattered wavefield of both longitudinal and transverse waves
even when the incident wave is purely longitudinal or purely
transverse.

The coefficients $b_{\ell m}^{+P}$ which determine the scattered wave
from the sphere at the origin are determined from the {\em total}
incident wave on that sphere, which consists of the incident plane
wave and the sum of the waves scattered from all the other spheres in
the plane. The latter, denoted by ${\bf u}'_{\rm sc}({\bf r})$, is
obtained from ${\bf u}_{\rm sc}({\bf r})$ by the removal of the term
corresponding to ${\bf R}_{n}={\bf 0}$ in Eq.(\ref{eq:multscat}).
${\bf u}'_{\rm sc}({\bf r})$ can be expanded into spherical waves
about the origin as follows
\begin{equation}
{\bf u}'_{\rm sc}({\bf r})=\sum_{\ell m}\left\{{\rm b}_{\ell m}^{'M}
j_{\ell}(q_tr){\bf X}_{\ell m} (\hat{{\bf r}})+{\rm b}_{\ell
m}^{'N}\frac{{\rm i}}{q_t}\nabla \times j_{\ell}(q_tr){\bf X}_{\ell
m} (\hat{{\bf r}})+{\rm b}_{\ell m}^{'L}\frac{1}{q_l}\nabla
\left[j_{\ell}(q_lr)Y_{\ell}^m (\hat{{\bf r}})\right] \right\}
\label{eq:scatprime}
\end{equation}
It can be shown (see Appendix~\ref{apxb}) that
\begin{equation}
{\rm b}_{\ell m}^{'P}=\sum_{P'\ell'm'} \Omega_{\ell m;\ell'm'}^{P P'}
{\rm b}_{\ell'm'}^{+P'}\;. \label{eq:bob}
\end{equation}
It is worth noting that the matrix elements of $\mathbf{\Omega}$
depend on the geometry (\ref{eq:Rn}) of the plane and, through
$q_{\nu}$, on the frequency, the mass density and the Lam\'e
coefficients of the medium surrounding the spheres of the plane; they
depend also on the reduced wavevector ${\mathbf{k}}_{\parallel}$ of
the incident wave; but they do not depend on the scattering
properties of the individual sphere.

The coefficients ${\rm b}_{\ell m}^{+P}$, which describe the
scattered wave from the sphere at the origin of the coordinates, are
given by
\begin{equation}
b^{+P}_{\ell m}=\sum_{P'\ell'm'} T_{\ell m;\ell'm'}^{PP'}\,\left({\rm
a}_{\ell'm'}^{0P'}+{\rm b}_{\ell'm'}^{'P'}\right). \label{eq:BTAB}
\end{equation}
The coefficients on the right-hand side of Eq.~(\ref{eq:BTAB})
describe the total wave incident on the sphere at the origin of
coordinates; ${\rm a}_{\ell m}^{0P}$ derive from the incident plane
wave given by Eq.~(\ref{eq:uinc}) via Eqs.~(\ref{eq:a0L}) and
(\ref{eq:a0P}), and ${\rm b}^{'P}_{\ell m}$ from the field defined by
Eq.~(\ref{eq:scatprime}). Combining Eqs.~(\ref{eq:bob}) and
(\ref{eq:BTAB}), we obtain
\begin{equation}
\sum_{P'\ell'm'} \left[\delta_{PP'}\delta_{\ell
\ell'}\delta_{mm'}-\sum_{P''\ell''m''} T_{\ell m;\ell''m''}^{PP''}\;
\Omega_{\ell'' m'';\ell'm'}^{P''P'} \right] {\rm
b}_{\ell'm'}^{+P'}=\sum_{P'\ell'm'} T_{\ell m;\ell'm'}^{PP'}\; {\rm
a}_{\ell'm'}^{0P'} \;. \label{eq:bsc}
\end{equation}
Eq.~(\ref{eq:bsc}) determines the coefficients $b_{\ell m}^{+P}$ of
the wave scattered from the plane of spheres, given by
Eq.~(\ref{eq:multscat}), in terms of the coefficients ${\rm a}_{\ell
m}^{0P}$ of the incident wave. According to Eqs.~(\ref{eq:a0L}) and
(\ref{eq:a0P}), we write the coefficients ${\rm a}_{\ell m}^{0P}$ of
the incident plane wave, defined by Eq.~(\ref{eq:uinc}), in the form
\begin{equation}
{\rm a}_{\ell m}^{0P}=\sum_{i'} A_{\ell m;i'}^{0P}({\bf K}_{{\bf
g}'\nu'}^{s'})\; [u_{in}]_{{\bf g}' i'}^{s'}\;, \label{eq:add1}
\end{equation}
where ${\bf A}_{\ell m}^{0P}$ are given by Eqs.~(\ref{eq:A0L}),
(\ref{eq:A0M}), and (\ref{eq:A0N}). Due to the linearity of
Eqs.~(\ref{eq:bsc}), the coefficients ${\rm b}_{\ell m}^{+P}$ can be
written as follows
\begin{equation}
{\rm b}_{\ell m}^{+P}=\sum_{i'} B_{\ell m;i'}^{+P}({\bf K}_{{\bf
g}'\nu'}^{s'})\; [u_{in}]_{{\bf g}' i'}^{s'}\;, \label{eq:add2}
\end{equation}
so that the system of Eqs.~(\ref{eq:bsc}) reduces to
\begin{eqnarray}
& & \sum_{P'\ell'm'} \left[\delta_{PP'}\delta_{\ell
\ell'}\delta_{mm'}-\sum_{P''\ell''m''} T_{\ell m;\ell''m''}^{PP''}\;
\Omega_{\ell'' m'';\ell'm'}^{P''P'} \right] B_{\ell'm';i'}^{+P'}({\bf
K}_{{\bf g}'\nu'}^{s'})=\sum_{P'\ell'm'} T_{\ell m;\ell'm'}^{PP'}\;
A_{\ell'm';i'}^{0P'} ({\bf K}_{{\bf g}'\nu'}^{s'})\,.\nonumber \\ & &
\label{eq:add3}
\end{eqnarray}
We remember that $i'$, $s'$, and ${\bf g}'$ are parameter values
characterizing the incident wave (we remember also that $\nu'$ is
determined by $i'$: $\nu'=l$ for $i'=1$ and $\nu'=t$ for $i'=2,3$).
Eqs.~(\ref{eq:add3}) (or, equivalently, Eqs.~(\ref{eq:bsc}))
constitute a system of infinitely many linear equations. It is solved
by introducing an angular momentum cut-off, $\ell_{\rm max}$,
truncating all angular momentum expansions to $\ell_{\rm max}$, thus
reducing the dimension of the system to $3\ell_{{\rm
max}}^2+6\ell_{{\rm max}}+1$. Moreover, by using the properties of
the matrix elements $\Omega_{\ell m;\ell'm'}^{PP'}$ given by
Eqs.~(\ref{eq:Osym}) of Appendix~\ref{apxb}, this system can be
reduced to two independent systems of $(3\ell_{{\rm
max}}^2+5\ell_{{\rm max}})/2$ and $(3\ell_{{\rm max}}^2+7\ell_{{\rm
max}}+2)/2$ linear equations, respectively.

Finally, the scattered wave given by Eq.~(\ref{eq:multscat}) can be
expressed as a sum of plane waves using the following identity
\begin{equation}
\sum_{{\mathbf{R}}_n} \exp({\rm i} {\mathbf{k}}_{\parallel} \cdot
{\mathbf{R}}_n)
h_{\ell}^+(q_{\nu}r_{n})Y_{\ell}^m(\hat{{\mathbf{r}}}_n)
=
\sum_{\mathbf{g}} \frac{2\pi (-{\rm i})^{\ell}}{q_{\nu} A_0
K_{{\mathbf{g}}\nu z}^+}
Y_{\ell}^m(\hat{{\mathbf{K}}}_{{\mathbf{g}}\nu}^{\pm}) \exp({\rm i}
{\mathbf{K}}_{{\mathbf{g}}\nu}^{\pm} \cdot \mathbf{r})
\label{eq:plasph}
\end{equation}
where $A_{0}$ denotes the area of the unit cell of the lattice given
by Eq.~(\ref{eq:Rn}). The plus (minus) sign on
${\mathbf{K}}_{{\mathbf{g}}\nu}$ must be used for $z>0$ ($z<0$). We
note that $K_{{\mathbf{g}}\nu z }^{\pm}$ can be real or imaginary. In
the latter case $\cos\theta_{{\mathbf{K}}_{{\mathbf{g}}\nu}^{\pm}}$
in the standard formulae for
$Y_{\ell}^m(\hat{{\mathbf{K}}}_{{\mathbf{g}}\nu}^{\pm})$ is replaced
by $K_{{\mathbf{g}}\nu z}^{\pm}/q_{\nu}$ (see text following
Eq.~(\ref{eq:e3})).

Using Eq.~(\ref{eq:plasph}) we can expand the scattered wave into a
series of longitudinal and transverse plane waves, as follows
\begin{equation}
{\mathbf{u}}_{\rm sc}^{s}({\mathbf{r}})= \sum_{{\mathbf{g}}} \sum_{P
\ell m} {\rm b}_{\ell m}^{+P} {\mathbf{\Delta}}_{\ell m}^{P}({\bf
K}_{{\bf g}\nu}^{s}) \exp({\rm i} {\mathbf{K}}_{{\mathbf{g}}\nu}^{s}
\cdot {\mathbf{r}}) \;, \label{eq:scplane1}
\end{equation}
where
\begin{mathletters}
\begin{eqnarray}
{\bf \Delta}_{\ell m}^{L}({\bf K}_{{\bf g}l}^s)=\frac{2\pi (-{\rm
i})^{\ell-1}}{q_l A_0 K_{{\bf g}lz}^+}Y_{\ell}^{m}({\bf K}_{{\bf
g}l}^s)\hat{{\bf e}}_1\;, \label{eq:DeltaL}
\end{eqnarray}
\begin{eqnarray}
&& {\bf \Delta}_{\ell m}^{M}({\bf K}_{{\bf g}t}^s) = \frac{2\pi
(-{\rm i})^{\ell}}{q_t A_0 K_{{\bf g}t z}^+\sqrt{\ell (\ell
+1)}}\nonumber
\\ && \times \biggl\{ \left[ \alpha_{\ell}^{-m}\cos \theta \;e^{{\rm
i}\phi}\;Y_{\ell}^{m-1}({\bf K}_{{\bf g}t}^s)-m\sin \theta \;
Y_{\ell}^{m}({\bf K}_{{\bf g}t}^s) + \alpha_{\ell}^{m} \cos \theta
\;e^{-{\rm i}\phi}\;Y_{\ell}^{m+1}({\bf K}_{{\bf g}t}^s)\right]
\hat{{\bf e}}_{2} \nonumber \\ && +{\rm i}
\left[\alpha_{\ell}^{-m}\;e^{{\rm i}\phi}\;Y_{\ell}^{m-1}({\bf
K}_{{\bf g}t}^s)-\alpha_{\ell}^{m}\;e^{-{\rm
i}\phi}\;Y_{\ell}^{m+1}({\bf K}_{{\bf g}t}^s)\right]\hat{{\bf
e}}_{3}\biggr\}\;, \label{eq:DeltaM} \\ && {\bf \Delta}_{\ell
m}^{N}({\bf K}_{{\bf g}t}^s)=\frac{2\pi ({-\rm i})^{\ell}}{q_t A_0
K_{{\bf g}t z}^+\sqrt{\ell (\ell +1)}}\biggl\{{\rm i}
\left[\alpha_{\ell}^{-m}\;e^{{\rm i}\phi}\;Y_{\ell}^{m-1}({\bf
K}_{{\bf g}t}^s)-\alpha_{\ell}^{m}\;e^{-{\rm
i}\phi}\;Y_{\ell}^{m+1}({\bf K}_{{\bf g}t}^s)\right]\hat{{\bf e}}_{2}
\nonumber \\&& -\left[\alpha_{\ell}^{-m}\cos \theta \;e^{{\rm
i}\phi}\;Y_{\ell}^{m-1}({\bf K}_{{\bf g}t}^s)-m\sin \theta\;
Y_{\ell}^{m}({\bf K}_{{\bf g}t}^s) +\alpha_{\ell}^{m} \cos \theta
\;e^{-{\rm i}\phi}\;Y_{\ell}^{m+1}({\bf K}_{{\bf
g}t}^s)\right]\hat{{\bf e}}_{3}\biggr\}\;,\label{eq:DeltaN}
\end{eqnarray}
\end{mathletters}
with $\theta$ and $\phi$ denoting the angular variables of ${\bf
K}_{{\bf g}t}^s$. Substituting ${\rm b}_{\ell m}^{+P}$ from
Eq.~(\ref{eq:add2}) into Eq.~(\ref{eq:scplane1}) we obtain
\begin{equation}
{\mathbf{u}}_{\rm sc}^{s}({\mathbf{r}})= \sum_{{\mathbf{g}}
i}\left[u_{sc}\right]_{{\bf g}i}^{s} \exp({\rm i}
{\mathbf{K}}_{{\mathbf{g}}\nu}^{s} \cdot {\mathbf{r})}\; \hat{{\bf
e}}_i \;, \label{eq:scplane}
\end{equation}
with
\begin{equation}
\left[u_{sc}\right]_{{\bf g}i}^{s}=\sum_{i'} \sum_{P \ell m}
\Delta_{\ell m;i}^{P}({\bf K}_{{\bf g}\nu}^{s}) B_{\ell
m;i'}^{+P}({\bf K}_{{\bf g}'\nu'}^{s'}) \left[u_{in}\right]_{{\bf
g}'i'}^{s'} \;, \label{eq:add4}
\end{equation}
where the superscript $s=+(-)$ holds for $z>0$ ($z<0$). We note that
the ${\mathbf{K}}_{{\mathbf{g}}\nu}^{s}$ in Eq.~(\ref{eq:scplane})
have the same frequency $\omega$ (the same wavenumber $q_{\nu}$) and
the same reduced wavevector ${\mathbf{k}}_{\parallel}$ as the
incident wave. We remember that for $i=1$, $\nu=l$ and, for a given
{\bf g}, $\hat{{\bf e}}_1$ is the radial unit vector along the
direction of ${\bf K}_{{\bf g}l}^s$. Similarly, for $i=2,3$, $\nu=t$
and, for given {\bf g}, $\hat{{\bf e}}_2,\hat{{\bf e}}_3$ are the
polar and azimuthal unit vectors, respectively, which are orthogonal
to ${\bf K}_{{\bf g}t}^s$. Eq.~(\ref{eq:scplane}) tells us that the
scattered wave consists, in general, of a number of diffracted beams,
of the same $\omega$ and ${\mathbf{k}}_{\parallel}$, corresponding to
different $\mathbf{g}$ vectors and polarization modes (longitudinal
or transverse). We note, however, that only beams for which
$K_{{\mathbf{g}}\nu z}^{s}$ is real constitute propagating waves. The
coefficients in Eq.~(\ref{eq:scplane}), given by Eq.~(\ref{eq:add4}),
are functions of the $ B_{\ell m;i'}^{+P}({\bf K}_{{\bf
g}'\nu'}^{s'})$ coefficients and through them depend on the incident
plane wave. These coefficients are to be evaluated for an incident
longitudinal ($i'=1$) or transverse ($i'=2,3$) plane wave, with a
wavevector ${\mathbf{K}}_{{\mathbf{g}}'\nu'}^{s'}$ given by
Eq.~(\ref{eq:Kg}), incident from the left ($s'=+$) or from the right
($s'=-$), with a displacement vector along the $i'$th direction of
magnitude equal to unity. In other words, $ B_{\ell m;i'}^{+P}({\bf
K}_{{\bf g}'\nu'}^{s'})$ are calculated from Eq.~(\ref{eq:add3}),
substituting $A_{\ell
m;i'}^{0P}({\mathbf{K}}_{{\mathbf{g}}'\nu'}^{s'})$, given by
Eqs.~(\ref{eq:A0L}), (\ref{eq:A0M}), and (\ref{eq:A0N}), on the
right-hand side of this equation. Obviously, when $i'=1$, only the
coefficients $A_{\ell m;i'}^{0L}({\mathbf{K}}_{{\mathbf{g}}'l}^{s'})$
are nonzero, and when $i'=2,3$ only $A_{\ell
m;i'}^{0M}({\mathbf{K}}_{{\mathbf{g}}'t}^{s'})$ and $A_{\ell
m;i'}^{0N}({\mathbf{K}}_{{\mathbf{g}}'t}^{s'})$ are nonzero.

\begin{figure}
\centering
 \epsfysize=6 cm \epsfbox{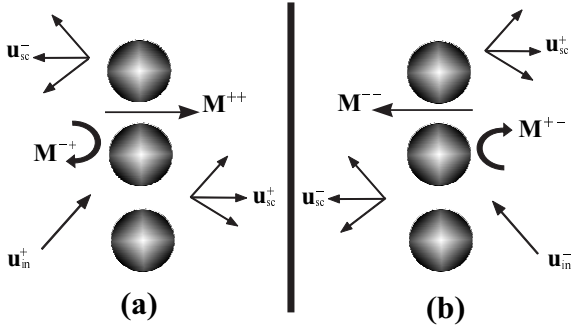}
\caption{Scattering of a plane elastic wave by a plane of spheres:
(a): the wave is incident from the left; (b): the wave is incident
from the right.} \label{fig1}
\end{figure}

Let us for the sake of clarity assume that a plane wave given by
Eq.~(\ref{eq:uinc}) is incident on the plane of spheres from the left
as in Fig.~\ref{fig1}a. Then the transmitted wave
(incident+scattered) on the right of the plane of spheres can be
written as
\begin{equation}
{\mathbf{u}}_{\rm tr}^+({\mathbf{r}})= \sum_{{\mathbf{g}} i}
\left[u_{\rm tr}\right]_{{\mathbf{g}} i}^+ \exp({\rm i}
{\mathbf{K}}_{{\mathbf{g}}\nu}^+ \cdot {\mathbf{r})} \hat{{\bf e}}_i
\,,\;\; z>0, \label{eq:transm}
\end{equation}
with
\begin{equation}
 \left[u_{\rm tr}\right]_{{\mathbf{g}} i}^+
=
 \left[u_{\rm in}\right]_{{\mathbf{g}}'i}^+
 \delta_{\mathbf{g}\mathbf{g}'}
+\left[u_{\rm sc}\right]_{{\mathbf{g}} i}^+
=
\sum_{i'}M_{{\mathbf{g}}i;{\mathbf{g}}'i'}^{++}
 \left[u_{\rm in}\right]_{{\mathbf{g}}' i'}^+
\label{eq:transampl}
\end{equation}
and the reflected wave as
\begin{equation}
{\mathbf{u}}_{\rm rf}^-({\mathbf{r}})= \sum_{{\mathbf{g}}
i}\left[u_{\rm rf}\right]_{{\mathbf{g}}i}^- \exp({\rm i}
{\mathbf{K}}_{{\mathbf{g}}\nu}^- \cdot {\mathbf{r})} \hat{{\bf
e}}_i\,,\;\; z<0, \label{eq:reflec}
\end{equation}
with
\begin{equation}
 \left[u_{\rm rf}\right]_{{\mathbf{g}} i}^-
=
\left[u_{\rm sc}\right]_{{\mathbf{g}} i}^-
=
\sum_{i'}M_{{\mathbf{g}}i;{\mathbf{g}}'i'}^{-+}
 \left[u_{\rm in}\right]_{{\mathbf{g}}'i'}^+\;.
\label{eq:refampl}
\end{equation}
Eqs.~(\ref{eq:add4}), (\ref{eq:transampl}) and (\ref{eq:refampl})
define the transmission ($M^{++}$) and reflection ($M^{-+}$) matrix
elements for a plane wave incident on the plane of spheres from the
left. Similarly, we can define the transmission matrix elements
$M_{{\mathbf{g}}i;{\mathbf{g}}'i'}^{--}$ and the reflection matrix
elements $M_{{\mathbf{g}}i;{\mathbf{g}}'i'}^{+-}$ for a plane wave
incident on the plane of spheres from the right (see
Fig.~\ref{fig1}b). We obtain
\begin{equation}
M_{{\mathbf{g}}i;{\mathbf{g}}'i'}^{s s'}= \delta_{s
s'}\delta_{\mathbf{g}\mathbf{g}'}\delta_{ii'} + \sum_{P \ell m}
\Delta_{\ell m;i}^{P}({\mathbf{K}}_{{\mathbf{g}}\nu}^{s})\;  B_{\ell
m;i'}^{+P}({\bf K}_{{\bf g}'\nu'}^{s'}) \;. \label{eq:mgg}
\end{equation}
One can show that the above matrix elements obey the following
symmetry relations
\begin{equation}
M_{{\mathbf{g}}i;{\mathbf{g}}'i'}^{-s-s'}=(-1)^{i+i'}\;
M_{{\mathbf{g}}i;{\mathbf{g}}'i'}^{s s'} . \label{eq:mgigi}
\end{equation}

\section{Scattering by a slab}
\label{slab} In what follows we need to evaluate the scattering
properties of a slab which by definition consists of a number of
layers (planes of spheres). For this purpose it is convenient to
express the plane waves on the left of a given plane of spheres with
respect to an origin, ${\mathbf{A}}_{l}$, on the left of the plane at
$-{\mathbf{d}}_{l}$ from its center, and the plane waves on the right
of this plane with respect to an origin, ${\mathbf{A}}_{r}$, on the
right of the plane at ${\mathbf{d}}_{r}$ from its center, i.e. a wave
on the left of the plane will be written as $\sum_{{\bf g}
i}u_{{\mathbf{g}}i}^{s}\exp \left[{\rm i}
{\mathbf{K}}_{{\mathbf{g}}\nu}^{s} \cdot
({\mathbf{r}}-{\mathbf{A}}_{l})\right]\hat{{\bf e}}_i$ and a wave on
the right of the plane will be written as $\sum_{{\bf g}
i}u_{{\mathbf{g}}i}^{s}\exp \left[{\rm i}
{\mathbf{K}}_{{\mathbf{g}}\nu}^{s} \cdot
({\mathbf{r}}-{\mathbf{A}}_{r})\right]\hat{{\bf e}}_i$. The
relationships between the amplitudes of the incident and of the
reflected and transmitted waves, when these are expressed with
respect to the above origins, follow directly from the corresponding
equations of Section~\ref{plane}. Accordingly, the amplitudes of
these waves are related through the ${\bf Q}$-matrix elements given
below
\begin{eqnarray}
Q_{{\mathbf{g}}i;{\mathbf{g}}'i'}^{{\rm I}}&=&
 M_{{\mathbf{g}}i;{\mathbf{g}}'i'}^{++}
\exp \left[{\rm i} (
 {\mathbf{K}}_{{\mathbf{g}}\nu}^{+}\cdot {\mathbf{d}}_{r}
+{\mathbf{K}}_{{\mathbf{g}}'\nu'}^{+}\cdot {\mathbf{d}}_{l} ) \right]
\nonumber\\ Q_{{\mathbf{g}}i;{\mathbf{g}}'i'}^{{\rm II}}&=&
 M_{{\mathbf{g}}i;{\mathbf{g}}'i'}^{+-}
\exp \left[{\rm i} (
 {\mathbf{K}}_{{\mathbf{g}}\nu}^{+}\cdot {\mathbf{d}}_{r}
-{\mathbf{K}}_{{\mathbf{g}}'\nu'}^{-}\cdot {\mathbf{d}}_{r} ) \right]
\nonumber\\ Q_{{\mathbf{g}}i;{\mathbf{g}}'i'}^{{\rm III}}&=&
 M_{{\mathbf{g}}i;{\mathbf{g}}'i'}^{-+}
\exp \left[-{\rm i} (
 {\mathbf{K}}_{{\mathbf{g}}\nu}^{-}\cdot {\mathbf{d}}_{l}
-{\mathbf{K}}_{{\mathbf{g}}'\nu'}^{+}\cdot {\mathbf{d}}_{l} ) \right]
\nonumber\\ Q_{{\mathbf{g}}i;{\mathbf{g}}'i'}^{{\rm IV}}&=&
 M_{{\mathbf{g}}i;{\mathbf{g}}'i'}^{--}
\exp \left[-{\rm i} (
 {\mathbf{K}}_{{\mathbf{g}}\nu}^{-}\cdot {\mathbf{d}}_{l}
+{\mathbf{K}}_{{\mathbf{g}}'\nu'}^{-}\cdot {\mathbf{d}}_{r} ) \right]
\;, \label{eq:qgg}
\end{eqnarray}
whose physical meaning is made obvious by their one-to-one
correspondence with the ${\bf M}$-matrix elements of
Section~\ref{plane}. From this point on, we shall write the above
matrices in compact form as: ${{\bf Q}}^{{\rm I}}$, ${{\bf Q}}^{{\rm
II}}$, ${{\bf Q}}^{{\rm III}}$ and ${{\bf Q}}^{{\rm IV}}$.

\begin{figure}
\centering
 \epsfysize=6cm \epsfbox{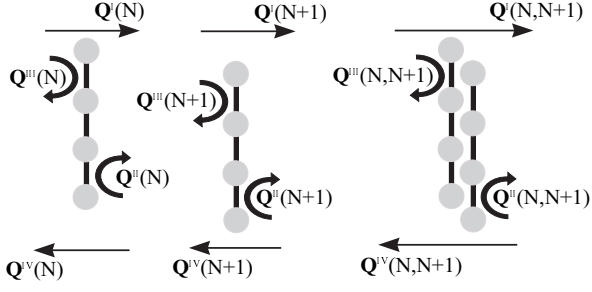}
\caption{The ${\bf Q}$ matrices for two successive layers are
obtained from those of the individual layers.} \label{fig2}
\end{figure}

\begin{figure}
\centering
 \epsfysize=6cm \epsfbox{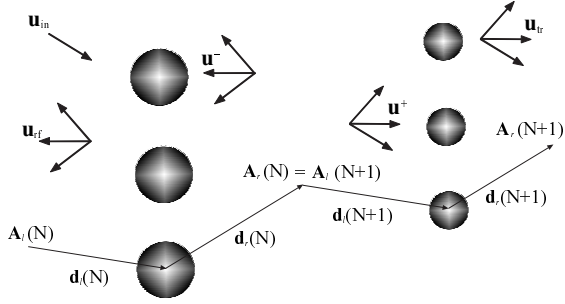}
\caption{Putting together a pair of planes of spheres.} \label{fig3}
\end{figure}

We obtain the transmission and reflection matrices for a pair of two
successive layers, $N$ and $N+1$, to be denoted by
${\mathbf{Q}}(N,N+1)$, by combining the matrices ${\mathbf{Q}}(N)$
and ${\mathbf{Q}}(N+1)$ of the two layers, as shown schematically in
Fig.~\ref{fig2}. One can easily show that
\begin{eqnarray}
{\bf Q}^{{\rm I}}(N,N+1)&=& {\bf Q}^{{\rm I}}(N+1) \left[
{\mathbf{I}}-{\bf Q}^{{\rm II}}(N){\bf Q}^{{\rm III}}(N+1) \right]
^{-1} {\bf Q}^{{\rm I}}(N) \nonumber \\ {\bf Q}^{{\rm II}}(N,N+1)&=&
{\bf Q}^{{\rm II}}(N+1)+{\bf Q}^{{\rm I}}(N+1) {\bf Q}^{{\rm II}}(N)
\nonumber \\ && \times \left[ {\mathbf{I}}-{\bf Q}^{{\rm
III}}(N+1){\bf Q}^{{\rm II}}(N) \right] ^{-1} {\bf Q}^{{\rm IV}}(N+1)
\nonumber \\ {\bf Q}^{{\rm III}}(N,N+1)&=& {\bf Q}^{{\rm
III}}(N)+{\bf Q}^{{\rm IV}}(N) {\bf Q}^{{\rm III}}(N+1) \nonumber \\
&& \times \left[ {\mathbf{I}}-{\bf Q}^{{\rm II}}(N){\bf Q}^{{\rm
III}}(N+1) \right] ^{-1} {\bf Q}^{{\rm I}}(N) \nonumber
\\ {\bf Q}^{{\rm IV}}(N,N+1)&=& {\bf Q}^{{\rm IV}}(N) \left[
{\mathbf{I}}-{\bf Q}^{{\rm III}}(N+1){\bf Q}^{{\rm II}}(N) \right]
^{-1} {\bf Q}^{{\rm IV}}(N+1)\,. \label{eq:qpair}
\end{eqnarray}
For example, knowing that
\begin{eqnarray}
\left[ {\mathbf{I}}-{\bf Q}^{{\rm II}}(N){\bf Q}^{{\rm III}}(N+1)
\right] ^{-1} & = & {\bf I}+{\bf Q}^{{\rm II}}(N){\bf Q}^{{\rm
III}}(N+1)\nonumber \\ & & +{\bf Q}^{{\rm II}}(N){\bf Q}^{{\rm
III}}(N+1){\bf Q}^{{\rm II}}(N){\bf Q}^{{\rm III}}(N+1)+\cdots\;,
\label{eq:proof1}
\end{eqnarray}
we can write the first of Eqs.~(\ref{eq:qpair}) as follows
\begin{eqnarray}
{\bf Q}^{{\rm I}}(N,N+1)&=& {\bf Q}^{{\rm I}}(N+1){\bf Q}^{{\rm
I}}(N)+{\bf Q}^{{\rm I}}(N+1){\bf Q}^{{\rm II}}(N){\bf Q}^{{\rm
III}}(N+1){\bf Q}^{{\rm I}}(N)\nonumber \\ & & + {\bf Q}^{{\rm
I}}(N+1){\bf Q}^{{\rm II}}(N){\bf Q}^{{\rm III}}(N+1){\bf Q}^{{\rm
II}}(N){\bf Q}^{{\rm III}}(N+1){\bf Q}^{{\rm I}}(N)+\cdots\;.
\label{eq:proof2}
\end{eqnarray}
The meaning of the terms is obvious. The first term signifies
transmission through the $N$th layer, followed by transmission
through the $(N+1)$th layer. The second term signifies transmission
through the $N$th layer, followed by reflection by the $(N+1)$th
layer, followed by reflection by the $N$th layer, followed by
transmission through the $(N+1)$th layer. The third and higher terms
can be interpreted in the same way: a wave incident from the left on
the pair of layers will be multiply reflected, any number of times,
between the layers before exiting the pair by transmission through
the second layer. In similar fashion one can understand the remaining
Eqs.~(\ref{eq:qpair}). All matrices refer of course to the same
$\omega$ and ${\bf k}_{\parallel}$. We remember that the waves on the
left (right) of the pair of layers are referred to an origin at
$-{\bf d}_{l}(N)$ ($+{\bf d}_{r}(N+1)$) from the center of the $N$th
($(N+1)$th) layer. The choice of ${\bf d}_{l}(N)$ and ${\bf
d}_{r}(N)$ is to some degree arbitrary, but it must be such that
${\bf A}_{r}(N)$ coincides with ${\bf A}_{l}(N+1)$, in accordance
with the definition of these quantities (see Fig.~\ref{fig3}).

It is obvious that by the same process we can obtain the transmission
and reflection matrices of three layers, by combining those of the
pair of layers with those of the third layer; and that we can in
similar fashion obtain the transmission and reflection matrices for a
slab consisting of any finite number of layers. In particular, having
calculated the $\mathbf{Q}$-matrix elements of a single layer, we can
obtain those of a slab of $N_{{\rm max}}=2^M$ identical layers by a
doubling-layer scheme as follows: we calculate the
$\mathbf{Q}$-matrix elements of two consecutive layers in the manner
described above, then using as units the $\mathbf{Q}$-matrix elements
of a pair of layers, we obtain those of four consecutive layers, and
in this way, by doubling the number of layers at each stage we obtain
the $\mathbf{Q}$-matrix elements of the slab.

In summary, for a plane wave of polarization $\nu'$,
$\sum_{i}\left[u_{\rm in}\right]_{{\mathbf{g}}'i}^{+} \exp\left[{\rm
i}{\mathbf{K}}_{\mathbf{g}'\nu'}^{+} \cdot
({\mathbf{r}}-{\mathbf{A}}_{L})\right]\hat{{\bf e}}_{i}$,\linebreak
incident on the slab from the left, we finally obtain a reflected
wave \linebreak $\sum_{{\bf g}i}\left[u_{\rm
rf}\right]_{{\mathbf{g}}i}^{-} \exp\left[ {\rm
i}{\mathbf{K}}_{\mathbf{g}\nu}^{-} \cdot
({\mathbf{r}}-{\mathbf{A}}_{L})\right]\hat{{\bf e}}_{i}$ on the left
of the slab and a transmitted wave $\sum_{{\bf g}i}\left[u_{\rm
tr}\right]_{{\mathbf{g}}i}^{+} \exp[ {\rm
i}{\mathbf{K}}_{\mathbf{g}\nu}^{+} \cdot
({\mathbf{r}}-{\mathbf{A}}_{R}) ]\hat{{\bf e}}_{i}$ on the right of
the slab, where ${\mathbf{A}}_{L}$ $({\mathbf{A}}_{R})$ is the
appropriate origin on the left (right) of the slab. We have
\begin{equation}
\left[u_{\rm tr}\right]_{{\mathbf{g}}i}^{+}
=
\sum_{i'} Q_{{\mathbf{g}}i;{\mathbf{g}}'i'}^{{\rm I}} \left[u_{\rm
in}\right]_{{\mathbf{g}}'i'}^{+} \label{eq:traref1}
\end{equation}
\begin{equation}
\left[u_{\rm rf}\right]_{{\mathbf{g}}i}^{-}
=
\sum_{i'} Q_{{\mathbf{g}}i;{\mathbf{g}}'i'}^{{\rm III}} \left[u_{\rm
in}\right]_{{\mathbf{g}}'i'}^{+} \label{eq:traref2}
\end{equation}
where the $\mathbf{Q}$-matrix elements are those of the slab. In the
present formulation of the problem we assume that the host material
between the spheres extends to the left and right of the slab to
infinity. However, the extension of the formalism to deal with
different materials on the left and right sides of the slab can be
easily effected by treating the interfaces as scattering elements
described by appropriate $\mathbf{Q}$-matrices, as in the case of
photonic crystals.~\cite{SKM92}

A transmitted beam (a plane wave with a real $K_{{\bf g}\nu z}^{+}$
component of the corresponding wavevector) carries with it an energy
flux density which, averaged over a time period $T=2\pi/\omega$,
gives (a formal proof of this formula can be obtained by applying the
standard definition of the Poynting vector $\mathbf{P}$ for elastic
waves, $P_i=-\sigma_{ik}\dot{u}_k$, to a plane wave~\cite{Auld})
\begin{equation}
{\bf P}_{{\bf g}\nu}^{\rm tr}=\frac{1}{2}\rho \omega c_{\nu}^2\left\{
\sum_i \left[u_{\rm tr}\right]_{{\bf g}i}^+\left(\left[u_{\rm
tr}\right]_{{\bf g}i}^+\right)^*\right\}{\bf K}_{{\bf g}\nu}^+\,.
\label{eq:Ptr}
\end{equation}
We recall that for a longitudinal wave ($\nu=l$) $i=1$, while for a
transverse wave ($\nu=t$) $i=2,3$; and $*$ denotes, as usual, complex
conjugation. We note that the quantity in braces in
Eq.~(\ref{eq:Ptr}) gives the square of the amplitude of the
displacement associated with the given plane wave. The transmitted
energy per unit area of the slab per unit time, associated with the
${\bf g},\nu$ beam, is given by the magnitude of the $z$ component,
$\left|P_{{\bf g}\nu z}^{\rm tr}\right|$, of ${\bf P}_{{\bf
g}\nu}^{\rm tr}$. A similar formula gives the energy flux associated
with any of the propagating reflected beams, or with the incident
wave. For the reflected beams we have
\begin{equation}
{\bf P}_{{\bf g}\nu}^{\rm rf}=\frac{1}{2}\rho \omega c_{\nu}^2\left\{
\sum_i \left[u_{\rm rf}\right]_{{\bf g}i}^-\left(\left[u_{\rm
rf}\right]_{{\bf g}i}^-\right)^*\right\}{\bf K}_{{\bf g}\nu}^-
\label{eq:Prf}
\end{equation}
And for the incident wave (of given ${\bf k}_{\|}+{\bf g}'$ and
polarized along the $i'$ direction) we obtain
\begin{equation}
{\bf P}_{{\bf g}' i'}^{\rm in}=\frac{1}{2}\rho \omega
c_{\nu'}^2\left\{ \left[u_{\rm in}\right]_{{\bf g}'
i'}^+\left(\left[u_{\rm in}\right]_{{\bf g}'
i'}^+\right)^*\right\}{\bf K}_{{\bf g}'\nu'}^+ \label{eq:Pin}
\end{equation}
The reflected energy per unit area of the slab per unit time
associated with the ${\bf g},\nu$ reflected beam is given by the
magnitude of the $z$ component, $\left|P_{{\bf g}\nu z}^{\rm
rf}\right|$, of ${\bf P}_{{\bf g}\nu}^{\rm rf}$ and the incident
energy per unit area of the slab per unit time is given by the
magnitude of the $z$ component, $\left|P_{{\bf g}' i' z}^{\rm
in}\right|$, of ${\bf P}_{{\bf g}' i'}^{\rm in}$. By definition the
reflectance is given by
\begin{equation}
{\mathcal{R}}(\omega,{\bf k}_{\|}+{\bf g}',i')=\frac{\sum_{{\bf
g}\nu} \left|P_{{\bf g}\nu z}^{\rm rf}\right|}{\left|P_{{\bf g}' i'
z}^{\rm in}\right|}= \frac {\sum_{{\mathbf{g}}i}c_{\nu}^2
\left[u_{\rm rf}\right]_{{\mathbf{g}}i}^{-} \left(\left[u_{\rm
rf}\right]_{{\mathbf{g}}i}^{-}\right)^{*} K_{{\mathbf{g}}\nu z}^{+}}
{ c_{\nu'}^2\left[u_{\rm in}\right]_{{\mathbf{g}}' i'}^{+}
\left(\left[u_{\rm in}\right]_{{\mathbf{g}}' i'}^{+}\right)^{*}
K_{{\mathbf{g}}'\nu'z}^{+}}\,, \label{eq:ref}
\end{equation}
and the transmittance by
\begin{equation}
{\mathcal{T}}(\omega,{\bf k}_{\|}+{\bf g}',i')=\frac{\sum_{{\bf
g}\nu} \left|P_{{\bf g}\nu z}^{\rm tr}\right|}{\left|P_{{\bf g}' i'
z}^{\rm in}\right|}= \frac {\sum_{{\mathbf{g}}i}c_{\nu}^2
\left[u_{\rm tr}\right]_{{\mathbf{g}}i}^{+} \left(\left[u_{\rm
tr}\right]_{{\mathbf{g}}i}^{+}\right)^{*} K_{{\mathbf{g}}\nu z}^{+}}
{ c_{\nu'}^2\left[u_{\rm in}\right]_{{\mathbf{g}}' i'}^{+}
\left(\left[u_{\rm in}\right]_{{\mathbf{g}}' i'}^{+}\right)^{*}
K_{{\mathbf{g}}'\nu'z}^{+}}\,, \label{eq:tra}
\end{equation}
where we have denoted explicitly the dependence of these coefficients
on the incident parameters. As long as there are no energy losses in
the slab, we have
\begin{equation}
{\mathcal{T}}+{\mathcal{R}}=1. \label{eq:conserve}
\end{equation}
\section{The complex band structure}
\label{band} We view the infinite crystal as a sequence of identical
layers parallel to the $xy$ plane, extending over all space (from $z
\rightarrow -\infty\ \mbox{to}\ z \rightarrow +\infty$). If
Eq.~(\ref{eq:Rn}) is the 2D space lattice for the layer, and
${\mathbf{a}}_{3}$ is a vector which takes us from a point in the
$N$th layer to an equivalent point in the $(N+1)$th layer, then
$\{{\mathbf{a}}_{1},{\mathbf{a}}_{2},{\mathbf{a}}_{3}\}$ is a set of
primitive vectors for the crystal.

In the region between the $N$th and the $(N+1)$th layers the
wavefield, of given $\omega$ and $\mathbf{k}_{\parallel}$, has the
form
\begin{equation}
{\mathbf{u}}({\mathbf{r}})= \sum_{{\mathbf{g}}i} \left\{
u_{{\mathbf{g}}i}^{+}(N) \exp\left[{\rm
i}{\mathbf{K}}_{{\mathbf{g}}\nu}^{+} \cdot
({\mathbf{r}}-{\mathbf{A}}_r(N))\right] + u_{{\mathbf{g}}i }^{-}(N)
\exp\left[{\rm i}{\mathbf{K}}_{{\mathbf{g}}\nu}^{-} \cdot
({\mathbf{r}}-{\mathbf{A}}_r(N))\right] \right\}\;{\hat{\bf e}}_i\;.
\label{eq:uNN+1}
\end{equation}

The coefficients $u_{{\mathbf{g}}i}^{s}(N)$ are related to the
 $u_{{\mathbf{g}}i}^{s}(N+1)$ coefficients through the scattering
properties of the $N$th layer. We have
\begin{eqnarray}
u_{{\mathbf{g}}i}^{-}(N) &=& \sum_{{\mathbf{g}}'i'}
Q_{{\mathbf{g}}i;{\mathbf{g}}'i'}^{{\rm IV}}
u_{{\mathbf{g}}'i'}^{-}(N+1) + \sum_{{\mathbf{g}}'i'}
Q_{{\mathbf{g}}i;{\mathbf{g}}'i'}^{{\rm III}}
u_{{\mathbf{g}}'i'}^{+}(N) \nonumber \\ u_{{\mathbf{g}}i}^{+}(N+1)
&=& \sum_{{\mathbf{g}}'i'} Q_{{\mathbf{g}}i;{\mathbf{g}}'i'}^{{\rm
I}} u_{{\mathbf{g}}'i'}^{+}(N) + \sum_{{\mathbf{g}}'i'}
Q_{{\mathbf{g}}i;{\mathbf{g}}'i'}^{{\rm II}}
u_{{\mathbf{g}}'i'}^{-}(N+1) \label{eq:egiegi}
\end{eqnarray}
where ${\bf Q}$ are the transmission/reflection matrices of the
layer.

A generalized Bloch wave, by definition, has the property
\begin{equation}
u_{{\mathbf{g}}i}^{s}(N+1)
=
\exp\left({\rm i}{\mathbf{k}} \cdot{\mathbf{a}}_{3}\right)
u_{{\mathbf{g}}i}^{s}(N)\,, \label{eq:blochwave}
\end{equation}
with
\begin{equation}
{\mathbf{k}}=\left({\mathbf{k}}_{\parallel},k_{z}(\omega,{\mathbf{k}}_{\parallel})
\right) \label{eq:bloch}
\end{equation}
where $k_{z}$ is, for a given ${\mathbf{k}}_{\parallel}$, a function
of $\omega$, to be determined.

We choose the reduced $\mathbf{k}$ zone of reciprocal space as
follows: $({\mathbf{k}}_{\parallel},k_{z})$ where
${\mathbf{k}}_{\parallel}=(k_{x},k_{y})$ extends over the SBZ of the
given crystallographic plane, and
$-\left|{\mathbf{b}}_{3}\right|/2<k_{z}
\leq\left|{\mathbf{b}}_{3}\right|/2$, where
${\mathbf{b}}_{3}\equiv2\pi{\mathbf{a}}_{1}\times{\mathbf{a}}_{2}/
{\mathbf{a}}_{1}\cdot({\mathbf{a}}_{2}\times{\mathbf{a}}_{3})=\hat{{\bf
e}}_z 2\pi /{\rm a}_{3z}$. The periodicity of the frequency band
structure parallel to the $xy$ plane follows from
Eq.~(\ref{eq:uNN+1}); for replacing ${\bf k}_{\|}$ by ${\bf
k}_{\|}+{\bf g}$ in this equation renames the coefficients without
changing the form of the wavefunction. Also, because the eigenvalues
of Eq.(\ref{eq:eig}) below are of the form $\exp\left({\rm i}{\bf
k}\cdot {\bf a}_{3}\right)$, values of $k_{z}$ differing by an
integral multiple of $\left|{\bf b}_{3}\right|$ correspond to the
same Bloch wave; which establishes the periodicity of the band
structure normal to the $xy$ plane. Substituting
Eq.(\ref{eq:blochwave}) into Eq.(\ref{eq:egiegi}) we obtain, after
some algebra (see Appendix~\ref{apxc}), the following system of
equations
\begin{eqnarray}
\left(
\begin{array}{cc}
{{\bf Q}}^{{\rm I}} & {{\bf Q}}^{{\rm II}}
\\
-\left[{{\bf Q}}^{{\rm IV}}\right]^{-1}{{\bf Q}}^{{\rm III}} {{\bf
Q}}^{{\rm I}} & \ \ \left[{{\bf Q}}^{{\rm IV}}\right]^{-1}
\left[{\mathbf{I}}-{{\bf Q}}^{{\rm III}}{{\bf Q}}^{{\rm II}}\right]
\end{array}
\right) \left(
\begin{array}{c}
{\mathbf{u}}^{+}(N) \\ {\mathbf{u}}^{-}(N+1)
\end{array}
\right) =\exp \left({\rm i}{\mathbf{k}}\cdot{\mathbf{a}}_{3}\right)
\left(
\begin{array}{c}
{\mathbf{u}}^{+}(N) \\ {\mathbf{u}}^{-}(N+1)
\end{array}
\right) \label{eq:eig}
\end{eqnarray}
where ${\mathbf{u}}^{\pm}$ are column matrices with elements:
$u_{{\bf g}_{1}1}^{\pm}$, $u_{{\bf g}_{1}2}^{\pm}$, $u_{{\bf
g}_{1}3}^{\pm}$, $u_{{\bf g}_{2}1}^{\pm}$, $u_{{\bf g}_{2}2}^{\pm}$,
$u_{{\bf g}_{2}3}^{\pm}$, $u_{{\bf g}_{3}1}^{\pm}$, $\cdots$. In
practice we keep $g_{{\rm max}}$ ${\mathbf{g}}$ vectors (those of the
smallest magnitude) in which case ${\bf u}^{\pm}$ are column matrices
with $3g_{\mathrm{max}}$ elements. The enumeration of the
$\mathbf{g}$ vectors implied in the above definition of ${\bf
u}^{\pm}$ is of course the same with the one implied in relation to
the ${\bf Q}$ matrices, each of which has $3g_{{\rm
max}}$$\times$$3g_{{\rm max}}$ elements; $\mathbf{I}$ is the
$3g_{{\rm max}}$$\times$$3g_{{\rm max}}$ unit matrix. For given ${\bf
k}_{\parallel}$ and $\omega$, we obtain $6g_{{\rm max}}$ eigenvalues
of $k_{z}$ from the eigenvalues of the $6g_{{\rm max}}\times 6g_{{\rm
max}}$ matrix on the left-hand side of Eq.(\ref{eq:eig}). The
eigenvalues $k_{z}(\omega;{\mathbf{k}}_{\parallel})$, looked upon as
functions of real $\omega$, define, for each
${\mathbf{k}}_{\parallel}$, $6g_{{\rm max}}$ lines in the complex
$k_{z}$ space. Taken together they constitute the complex band
structure of the infinite crystal associated with the given
crystallographic plane. A line of given $\mathbf{k}_{\parallel}$ may
be real (in the sense that $k_{z}$ is real) over certain frequency
regions, and be complex (in the sense that $k_{z}$ is complex) for
$\omega$ outside these regions. It turns out that for given
$\mathbf{k}_{\parallel}$ and $\omega$, out of the $6g_{{\rm max}}$
eigenvalues of $k_{z}(\omega;{\mathbf{k}}_{\parallel})$ none or, at
best, a few are real; the eigensolutions of Eq.(\ref{eq:eig})
corresponding to them, represent propagating modes of the elastic
field in the given infinite crystal. The remaining eigenvalues of
$k_{z}(\omega;\mathbf{k}_{\parallel})$ are complex and the
corresponding eigensolutions represent evanescent waves. These have
an amplitude which increases exponentially in the positive or
negative $z$ direction and, unlike the propagating waves, do not
exist as physical entities in the infinite crystal. However, they are
an essential part of the physical solutions of the elastic field in a
semi-infinite crystal (extending from $z \rightarrow -\infty$ to
$z=0$) or in a slab of finite thickness. A region of frequency where
propagating waves do not exist for given ${\mathbf{k}}_{\parallel}$
constitutes a frequency gap of the elastic field for the given
${\mathbf{k}}_{\parallel}$. If over a frequency region no propagating
wave exists whatever the value of $\mathbf{k}_{\parallel}$, then this
region constitutes an (absolute) frequency gap.

Finally, it is worth noting that, when there is a plane of mirror
symmetry associated with the surface under consideration, the
eigensolutions (Bloch waves) of Eq.(\ref{eq:eig}) appear in pairs:
 $k_{z}(\omega;{\mathbf{k}}_{\parallel})$ and
$-k_{z}(\omega;{\mathbf{k}}_{\parallel})$.
\section{An example}
\label{applic} 
\begin{figure}
\centering
 \epsfysize=6 cm \epsfbox{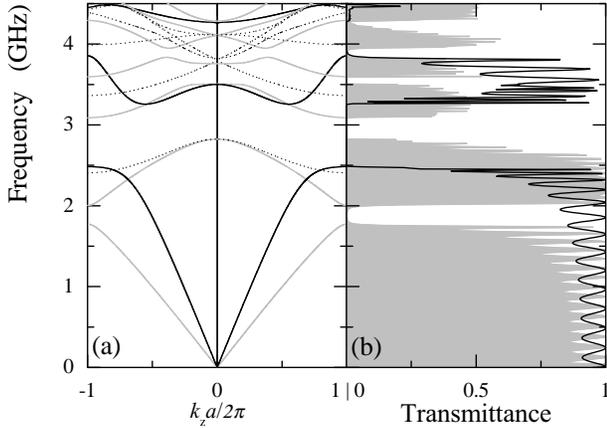}
\caption{The phononic band structure at the center of the SBZ of a
(001) surface of an fcc crystal of silica spheres in ice (a); and the
corresponding transmittance curve of a slab of 16 layers parallel to
the same surface (b). The lattice constant is 1~$\mu$m and the radius
of the spheres is 0.25~$\mu$m. In (a) the black lines represent
longitudinal modes (in the sense defined in the text), the grey lines
transverse modes, and the dotted lines are deaf bands.
Correspondingly in (b) the solid line shows the transmittance for
longitudinal incident elastic waves; and the shaded curve that for
transverse incident waves.} \label{fig4}
\end{figure}

We demonstrate the applicability of our method by
applying it to a specific example, which has also been considered by
Sprik and Wegdam:~\cite{Sprik98} a system of silica spheres of radius
$S=0.25~\mu$m centered on the sites of an fcc lattice with a lattice
constant of 1~$\mu$m; the host material being ice. The relevant
parameters are, for silica: $\rho=2200$~Kgm$^{-3}$,
$c_l=5970$~ms$^{-1}$, $c_t=3760$~ms$^{-1}$, and for ice (at
$-16$~$^{\circ}$C): $\rho=940$~Kgm$^{-3}$, $c_l=3830$~ms$^{-1}$,
$c_t=1840$~ms$^{-1}$. We view the crystal as a succcession of planes
of spheres parallel to the (001) direction of the fcc lattice.
Fig.~\ref{fig4} shows the frequency band structure normal to the
(001) plane (${\bf k}_{\|}={\bf 0}$) and the corresponding
transmission spectrum for both longitudinal and transverse waves
incident normally on a slab of the above crystal consisting of 16
layers.

To begin with, we compare the band structure normal to the (001)
plane with the results of Sprik and Wegdam ~\cite{Sprik98} obtained
by the plane-wave method (using 343 plane waves) with an accuracy, as
stated by the above authors, of a few percent. Our results obtained
with an angular momentum cutoff $\ell_{max}=4$ and 13 ${\bf g}$
vectors are converged within an accuracy of $10^{-3}$, and they agree
with those of Sprik and Wegdam ~\cite{Sprik98} within the stated
accuracy of their results. Furthermore, the evaluation of the
transmission coefficient, easily obtainable by our method but not
possible by the plane-wave method, confirms the validity of the
band-structure calculation. The transmittance curves are shown in
Fig.~\ref{fig4} by the shaded curve for the transverse waves and by
the solid line for the longitudinal waves. The oscillations in the
transmittance curves are due to multiple reflections at the edges of
the slab (Fabry-Perot like oscillations). As expected, for
frequencies within a gap the corresponding transmission coefficient
vanishes.

The eigenmodes of the phononic crystal are strictly speaking always
hybrid, having a longitudinal and a transverse component. However
along symmetry directions the following situation may arise. We
consider for simplicity the normal modes corresponding to ${\bf
k}_{\|}={\bf 0}$ (the direction normal to the surface). In this case
the component of the field associated with the ${\bf g}={\bf 0}$ beam
is either longitudinal or transverse; the field associated with the
${\bf g}\neq{\bf 0}$ components need not be of the same type.
However, only the ${\bf g}={\bf 0}$ component couples to the external
field (incident, reflected and transmitted waves), if $\left({\bf
k}_{\|}+{\bf g}\right)^2>q_{\nu}^2$ for ${\bf g}\neq{\bf 0}$.
Therefore an incident longitudinal or transverse wave will excite a
mode (or modes) in the interior of the crystal with a ${\bf g}={\bf
0}$ component of the same type. As long as the amplitude of the ${\bf
g}={\bf 0}$ component of these modes is much greater than those of
the ${\bf g}\neq{\bf 0}$ components, which is the case in the example
we have considered, the transmitted and reflected waves will be of
the same type as the incident wave, but this need not be the case in
general. Our results shown in Fig.~\ref{fig4} are termed longitudinal
or transverse in the above sense. For waves incident at an angle on
the surface of the slab (${\bf k}_{\|}\neq{\bf 0}$) the above
distinction between longitudinal and transverse waves no longer
applies (the ${\bf g}={\bf 0}$ component of the elastic field inside
the crystal is a hybrid one) and therefore an incident wave of a
specific type (longitudinal or transverse) will give rise to
reflected and transmitted waves of a mixed type.

\begin{figure}
\centering
 \epsfysize=6 cm \epsfbox{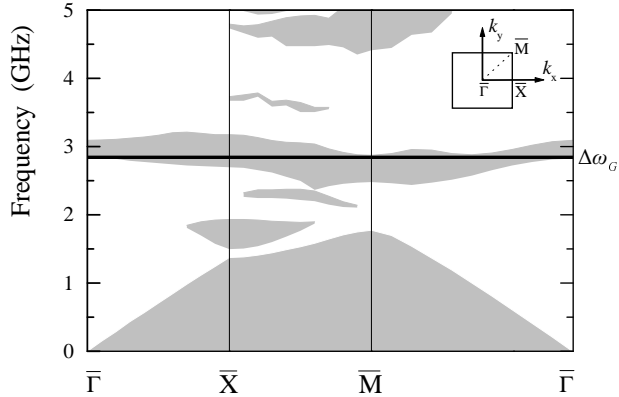}
\caption{Projection of the frequency band structure on the SBZ of the
(001) surface of the fcc phononic crystal described in the caption of
Fig.~\ref{fig4}. The shaded areas show the frequency gaps in the
considered frequency region. The inset shows the SBZ of the (001)
surface.} \label{fig5}
\end{figure}

In Fig.~\ref{fig5} we show the projection of the frequency band
structure on the SBZ of the (001) plane along its symmetry lines.
This is obtained, for a given ${\bf k}_{\|}$, as follows: the regions
of $\omega$ for which there are no propagating states in the infinite
crystal (the corresponding values of all $k_z\left(\omega,{\bf
k}_{\|}\right)$ are complex), are shown shaded, against the white
areas which correspond to regions over which propagating states do
exist (for a given $\omega$ there is at least one solution
corresponding to $k_z\left(\omega,{\bf k}_{\|}\right)$ real). We note
the existence of a narrow absolute gap, denoted by $\Delta \omega
_G$, extending from 2.82~GHz to 2.89~GHz. An absolute gap at
approximately the same frequency and of approximately the same width
was found by Sprik and Wegdam.~\cite{Sprik98}

A considerable number of bands of longitudinal and transverse waves
exist above the absolute gap (see Fig.~\ref{fig4}). Below this gap we
have two bands of transverse waves, which are doubly degenerate,
extending from 0~GHz to 1.77~GHz and from 2.01~GHz to 2.82~GHz, with
a gap in between. On the other hand, a non-degenerate band of
longitudinal waves extends from 0~GHz to 2.48~GHz. In addition to
these bands, we find a non-degenerate band, extending from 2.40~GHz
to 2.82~GHz, for which the ${\bf g}={\bf 0}$ component of the
corresponding eigenmodes of the elastic field vanishes. Because the
${\bf g}={\bf 0}$ beam is the only one which matches (couples with) a
propagating wave outside the crystal, an internal mode with a
vanishing ${\bf g}={\bf 0}$ component is not excited by the incident
wave. Therefore, if this were the only band over the stated frequency
region, the wave would be totally reflected. However, in our example
transmission through the slab in the frequency range of this deaf
band occurs, because other bands with non-vanishing ${\bf g}={\bf 0}$
components exist in the same frequency region. We note that analogous
deaf bands are known to exist in photonic crystals.~\cite{K98}

The long wavelength limit ($k_z \rightarrow 0$) is represented by the
linear segments of the dispersion curves, the slopes of which
determine the propagation velocities of longitudinal and transverse
waves ($\overline{c}_l=3893$~ms$^{-1}$,
$\overline{c}_t=2033$~ms$^{-1}$) in a corresponding effective medium.
\section{Conclusion}
\label{conclusion} We have shown that for a system of non-overlapping
elastic spheres arranged periodically in a host medium of different
elastic coefficients one can, using the formalism of the present
paper, calculate accurately and efficiently the phonon spectrum of
the infinite crystal and, also, the transmission, reflection, and
absorption coefficients of elastic waves incident on a slab of the
material of finite thickness.
\appendix
\section{}
\label{apxa} The nonzero elements of the {\bf T} matrix for a solid
sphere in a solid host are
\begin{eqnarray}
T_{\ell m;\ell' m'}^{MM}= && \frac{(\rho_s z^2_t/\rho
x^2_t)j_{\ell}(z_t)
[x_tj'_{\ell}(x_t)-j_{\ell}(x_t)]-j_{\ell}(x_t)[z_tj'_{\ell}(z_t)-j_{\ell}(z_t)]}
{j_{\ell}(x_t)[z_t{h_{\ell}^{+}}'(z_t)-h_{\ell}^+(z_t)]- (\rho_s
z^2_t/\rho
x^2_t)h_{\ell}^+(z_t)[x_tj'_{\ell}(x_t)-j_{\ell}(x_t)]}\delta_{\ell
\ell'} \delta_{mm'} \;,\; \ell,\ell'\geq 1,\nonumber \\ T_{\ell
m;\ell' m'}^{NN}= && \frac{W_{\ell}^{NN}}{D_{\ell}}\delta_{\ell
\ell'} \delta_{mm'} \;,\; \ell,\ell'\geq 1,\nonumber \\ T_{\ell
m;\ell' m'}^{NL}= && (z_t/z_l)\frac{W_{\ell}^{NL}\sqrt{\ell(\ell+1)}}
{D_{\ell}}\delta_{\ell \ell'} \delta_{mm'} \;,\; \ell\geq
1,\;\ell'\geq0, \nonumber \\ T_{\ell m;\ell'm' }^{LN}= &&
(z_l/z_t)\frac{W_{\ell}^{LN}}
{D_{\ell}\sqrt{\ell(\ell+1)}}\delta_{\ell \ell'} \delta_{mm'} \;,\;
\ell \geq 0,\; \ell' \geq 1\;,\nonumber \\ T_{\ell m;\ell' m' }^{LL}=
&& \frac{W_{\ell}^{LL}}{D_{\ell}}\delta_{\ell \ell'} \delta_{mm'}
\;,\; \ell,\ell'\geq 0\;, \label{eq:Tmatrix}
\end{eqnarray}
with $z_{\nu}=Sq_{\nu}$, $x_{\nu}=Sq_{s\nu}$ and $\nu=l,t$. The
elements of the $4\times4$ determinant $D_{\ell}$ are given by
\begin{equation}
\begin{array}{l}
d_{11}=z_t{h_{\ell}^{+}}'(z_t)+h_{\ell}^+(z_t)\\
d_{21}=\ell(\ell+1)h_{\ell}^+(z_t)\\
d_{31}=\left[\ell(\ell+1)-z_t^2/2-1\right]h_{\ell}^+(z_t)-z_t
{h_{\ell}^{+}}'(z_t)\\
d_{41}=\ell(\ell+1)\left[z_t{h_{\ell}^{+}}'(z_t)-h_{\ell}^+(z_t)\right]\\
d_{12}=h_{\ell}^+(z_l)\\ d_{22}=z_l{h_{\ell}^{+}}'(z_l)\\
d_{32}=z_l{h_{\ell}^{+}}'(z_l)-h_{\ell}^+(z_l)\\
d_{42}=\left[\ell(\ell+1)-z_t^2/2\right]h_{\ell}^+(z_l)-2z_l
{h_{\ell}^{+}}'(z_l)\\ d_{13}=x_tj'_{\ell}(x_{t})+j_{\ell}(x_t)\\
d_{23}=\ell(\ell+1)j_{\ell}(x_t)\\ d_{33}=(\rho_s z_t^2/\rho x_t^2)
\left\{\left[\ell(\ell+1)-x_{t}^2/2-1\right]j_{\ell}(x_t)-x_t
j'_{\ell}(x_t)\right\}\\ d_{43}=(\rho_s z_t^2/\rho x_t^2)
\ell(\ell+1)\left[x_{t}j'_{\ell}(x_t)-j_{\ell}(x_t)\right]\\
d_{14}=j_{\ell}(x_l)\\ d_{24}=x_lj'_{\ell}(x_l)\\ d_{34}=(\rho_s
z_t^2/\rho x_t^2)\left[ x_lj'_{\ell}(x_l)-j_{\ell}(x_l)\right]\\
d_{44}=(\rho_s z_t^2/\rho x_t^2)
\left\{\left[\ell(\ell+1)-x_t^2/2\right]j_{\ell}(x_l)-2x_lj'_{\ell}(x_l)\right\}\;,
\end{array}
\label{eq:dij}
\end{equation}
where $j'_{\ell}$ and ${h_{\ell}^{+}}'$ denote the first derivatives
of the spherical Bessel and Hankel functions, respectively.
$W_{\ell}^{PP'}$ are given by the following determinants
\begin{eqnarray}
W_{\ell}^{NN} &=& -\left|
\begin{array}{llll}
d_{1}^N & d_{12} & d_{13} & d_{14}\\ d_{2}^N & d_{22} & d_{23} &
d_{24}\\ d_{3}^N & d_{32} & d_{33} & d_{34}\\ d_{4}^N & d_{42} &
d_{43} & d_{44}
\end{array}
\right|\,,\;\;\;\;\; W_{\ell}^{NL}= \;\;\,\left|
\begin{array}{llll}
d_{1}^L & d_{12} & d_{13} & d_{14}\\ d_{2}^L & d_{22} & d_{23} &
d_{24}\\ d_{3}^L & d_{32} & d_{33} & d_{34}\\ d_{4}^L & d_{42} &
d_{43} & d_{44}
\end{array}
\right|\,, \nonumber \\ \nonumber \\ \nonumber \\ W_{\ell}^{LN} &=&
\;\;\,\,\left|
\begin{array}{llll}
d_{11} & d_{1}^N & d_{13} & d_{14}\\ d_{21} & d_{2}^N & d_{23} &
d_{24}\\ d_{31} & d_{3}^N & d_{33} & d_{34}\\ d_{41} & d_{4}^N &
d_{43} & d_{44}
\end{array}
\right|\,,\;\;\;\; W_{\ell}^{LL}= \,-\left|
\begin{array}{llll}
d_{11} & d_{1}^L & d_{13} & d_{14}\\ d_{21} & d_{2}^L & d_{23} &
d_{24}\\ d_{31} & d_{3}^L & d_{33} & d_{34}\\ d_{41} & d_{4}^L &
d_{43} & d_{44}
\end{array}
\right|\,, \label{eq:W}
\end{eqnarray}
where
\begin{equation}
\begin{array}{l}
d_{1}^N=z_tj'_{\ell}(z_t)+j_{\ell}(z_t)\\
d_{2}^N=\ell(\ell+1)j_{\ell}(z_t)\\
d_{3}^N=\left[\ell(\ell+1)-z_t^2/2-1\right]j_{\ell}(z_t)-z_t
j'_{\ell}(z_t)\\
d_{4}^N=\ell(\ell+1)\left[z_tj'_{\ell}(z_t)-j_{\ell}(z_t)\right]\;,
\end{array}
\label{eq:dN}
\end{equation}
and
\begin{equation}
\begin{array}{l}
d_{1}^L=j_{\ell}(z_l)\\ d_{2}^L=z_lj'_{\ell}(z_l)\\
d_{3}^L=z_lj'_{\ell}(z_l)-j_{\ell}(z_l)\\
d_{4}^L=\left[\ell(\ell+1)-z_t^2/2\right]j_{\ell}(z_l)-2z_l
j'_{\ell}(z_l)\;.
\end{array}
\label{eq:dp}
\end{equation}
\section{}
\label{apxb} A longitudinal (transverse) spherical wave about ${\bf
R}_n \neq {\bf 0}$ remains a longitudinal (transverse) wave when
expanded about the origin of coordinates (${\bf R}_n ={\bf 0}$), and
therefore the matrix elements of ${\bf {\Omega}}$ defined by
Eq.~(\ref{eq:bob}) are obtained independently for longitudinal and
transverse waves.

For the transverse waves the evaluation of these elements proceeds as
in the case of the electromagnetic (EM) field described in
Ref.~\onlinecite{Mod87}. We note that the $M$ transverse elastic wave
corresponds to the $H$ component of the electric field of the EM wave
and the $N$ transverse elastic wave corresponds to the $E$ component
of the electric field of the EM wave (compare Eqs.~(15) and (16) of
Ref.~\onlinecite{Mod87} with Eqs.~(\ref{eq:multscat}) and
(\ref{eq:scatprime}) of the present article). Therefore, the
$\Omega^{PP'}$-matrix elements with $P=M,N$ and $P'=M,N$ can be taken
directly from Ref.~\onlinecite{Mod87}. Taking into account the fact
that the expansion coefficients in Eq.~(\ref{eq:multscat}) above and
those in Eq.~(15) of Ref.~\onlinecite{Mod87} are multiplied by
different constants, one readily obtains
\begin{eqnarray}
\Omega_{\ell m;\ell'm'}^{MM}&=& \frac{2\alpha_{\ell}^{-m}
\alpha_{\ell'}^{-m'}Z_{\ell' m'-1}^{\ell m-1}(q_t) +
mm'Z_{\ell'm'}^{\ell
m}(q_t)+2\alpha_{\ell}^m\alpha_{\ell'}^{m'}Z_{\ell'm'+1}^{\ell
m+1}(q_t)}
{\left[\ell(\ell+1)\ell'(\ell'+1)\right]^{1/2}},\;\;\ell,\ell' \geq1,
\nonumber
\\ \Omega_{\ell m;\ell'm'}^{NN}&=& \Omega_{\ell m;\ell'm'}^{MM},\;\;\ell,\ell'
\geq1, \label{eq:Omega1}
\end{eqnarray}
\begin{eqnarray}
\Omega_{\ell m;\ell' m'}^{MN} &=& -\Omega_{\ell m;\ell' m'}^{NM}= (2
\ell+1)\left[\ell(\ell+1)\ell'(\ell'+1)\right]^{-1/2}\nonumber
\\ & & \times
\biggl\{(8\pi/3)^{1/2}(-1)^{m}\alpha_{\ell'}^{m'}Z_{\ell' m'
+1}^{\ell-1 m+1}(q_t)B_{\ell-1,m+1}(1,-1;\ell m)\nonumber
\\ & & -(8\pi/3)^{1/2}(-1)^{m}\alpha_{\ell'}^{-m'}Z_{\ell' m'
-1}^{\ell-1 m-1}(q_t)B_{\ell-1,m-1}(1,1;\ell m)\nonumber
\\ & & +m'Z_{\ell' m'}^{\ell-1
m}(q_t)[(\ell+m)(\ell-m)/(2\ell-1)(2\ell+1)]^{1/2}\biggr\},\;\;\ell,\ell'
\geq1, \label{eq:Omega2}
\end{eqnarray}
where
\begin{equation}
Z_{\ell m}^{\ell'm'}(q_t)\equiv\sum_{{\bf R}_n\neq {\bf 0}} \exp({\rm
i} {\bf k}_{\|}\cdot {\bf R}_n) G_{\ell m;\ell'm'}(-{\bf R}_n;q_t)\,,
\label{eq:zlm}
\end{equation}
\begin{eqnarray}
G_{\ell m;\ell''m''}(-{\bf R}_n;q_t) &\equiv& 4\pi \sum_{\ell'm'}
(-1)^{(\ell-\ell'-\ell'')/2} (-1)^{m'+m''}B_{\ell
m}(\ell'm';\ell''m'')\nonumber \\ & & \times h_{\ell'}^{+}(q_tR_n)
Y_{\ell'}^{-m'}(-\hat{{\bf R}}_n)\,, \label{eq:glm}
\end{eqnarray}
\begin{equation}
B_{\ell m}(\ell'm';\ell''m'')\equiv \int {\rm d}\hat{{\bf r}}\;
     Y_{\ell}^{m}(\hat{{\bf r}})Y_{\ell'}^{m'}(\hat{{\bf r}})
     Y_{\ell''}^{-m''}(\hat{{\bf r}})\,.
     \label{eq:BInt}
     \end{equation}
The expression for $\Omega^{MN}$ can be simplified further by the
evaluation of the $B_{\ell m}$ coefficients defined by
Eq.~(\ref{eq:BInt}). Using standard formulas (see, e.g,
Ref.~\onlinecite{Rose}) one finally obtains
\begin{eqnarray}
\Omega_{\ell m;\ell'm'}^{MN}&=& (2\ell+1)\frac{-2\alpha_{\ell'}^{-m'}
\gamma_{\ell}^mZ_{\ell'm'-1}^{\ell-1m-1}(q_t)
+m'\zeta_{\ell}^mZ_{\ell'm'}^{\ell-1m}(q_t)+2\alpha_{\ell'}^{m'}\gamma_{\ell}^{-m}
Z_{\ell'm'+1}^{\ell-1m+1}(q_t)}{\left[\ell(\ell+1)\ell'(\ell'+1)\right]^{1/2}}
,\;\;\ell,\ell' \geq1, \nonumber \\ \Omega_{\ell m;\ell'm'}^{NM}&=&
-\Omega_{\ell m;\ell'm'}^{MN},\;\;\ell,\ell' \geq1,
\label{eq:strconst}
\end{eqnarray}
where
\begin{eqnarray}
&&
\gamma^m_{\ell}=\frac{1}{2}[(\ell+m)(\ell+m-1)]^{1/2}/[(2\ell-1)(2\ell+1)]^
{1/2}\;\;{\rm and} \nonumber
\\ && \zeta^m_{\ell}=[(\ell+m)(\ell-m)]^{1/2}/[(2\ell-1)(2\ell+1)]^{1/2}\;.
\label{eq:betgamlm}
\end{eqnarray}
The derivation of the above formulae is based on the following
relation~\cite{Mod87}
\begin{eqnarray}
h_{\ell}^{+}(qr_n)Y_{\ell}^{m}(\hat{{\bf r}}_n)=\sum_{\ell'm'}
G_{\ell m;\ell'm'}(-{\bf R}_n;q)j_{\ell'}(qr)Y_{\ell'}^{m'}(\hat{{\bf
r}})\,, \label{eq:hYtoGjY}
\end{eqnarray}
which expresses a scalar spherical wave about ${\bf R}_n\neq {\bf
0}$, as a sum of spherical waves about the origin ({${\bf R}_n={\bf
0}$). The longitudinal wave, described by the third term of
Eq.~(\ref{eq:multscat}), is obtained by multiplying
Eq.~(\ref{eq:hYtoGjY}) with $\exp({\rm i}{\bf k}_{\|}\cdot{\bf R}_n)$
and summing over all ${\bf R}_n\neq {\bf 0}$, which immediately tells
us that
\begin{equation}
\Omega_{\ell m;\ell'm'}^{LL}=Z_{\ell'm'}^{\ell m}(q_l),\;\;\ell,\ell'
\geq 0, \label{eq:OmegaLL}
\end{equation}

The evaluation of the matrices $\mbox{\boldmath $\Omega$}$ involves
the evaluation of the matrix ${\bf Z}$ which is a well known quantity
in the theory of low-energy electron diffraction (LEED) and a
computer program for its evaluation is already available in the
literature.~\cite{Pendry} Further calculation is made simpler by the
following property of $Z_{\ell m}^{\ell'm'}(q_{\nu})$~\cite{Pendry}
\begin{equation}
Z_{\ell m}^{\ell'm'}(q_{\nu})=0,\;\; {\rm unless}\;\;
\ell+m+\ell'+m':\;\; {\rm even}. \label{eq:zsym}
\end{equation}
It follows from Eq.~(\ref{eq:zsym}) that
\begin{eqnarray}
\Omega_{\ell m;\ell'm'}^{MM} &=& \Omega_{\ell
m;\ell'm'}^{NN}=\Omega_{\ell m;\ell'm'}^{LL}=0,\;\;  {\rm unless}\;\;
\ell+m+\ell'+m':\;\; {\rm even},\nonumber \\ \Omega_{\ell
m;\ell'm'}^{MN} &=& \Omega_{\ell m;\ell'm'}^{NM}=0,\;\; {\rm
unless}\;\; \ell+m+\ell'+m':\;\; {\rm odd}. \label{eq:Osym}
\end{eqnarray}
\section{}
\label{apxc} Eq.~(\ref{eq:eig}), initially derived by McRae for the
case of electron scattering by atomic layers~\cite{McRae} can be
proven as follows. We replace the quantities on the left of
Eqs.~(\ref{eq:egiegi}) with the aid of Eq.~(\ref{eq:blochwave}) to
obtain
\begin{eqnarray}
\exp \left({\rm i}{\mathbf{k}}\cdot{\mathbf{a}}_{3}\right) \left(
\begin{array}{cc}
{\bf I} & {\bf 0}
\\
{{\bf Q}}^{{\rm III}} & {{\bf Q}}^{{\rm IV}}
\end{array}
\right) \left(
\begin{array}{c}
{\mathbf{u}}^{+}(N) \\ {\mathbf{u}}^{-}(N+1)
\end{array}
\right)= \left( \begin{array}{cc} {{\bf Q}}^{{\rm I}} & {{\bf
Q}}^{{\rm II}}
\\
{\bf 0} & {\bf I}
\end{array}
\right) \left(
\begin{array}{c}
{\mathbf{u}}^{+}(N) \\ {\mathbf{u}}^{-}(N+1)
\end{array}
\right)\;.
 \label{eq:McRae1}
\end{eqnarray}
The inverse of the matrix on the left of Eq.~(\ref{eq:McRae1}) is
given by
\begin{eqnarray}
\left(
\begin{array}{cc}
{\bf I} & {\bf 0}
\\
-\left[{{\bf Q}}^{{\rm IV}}\right]^{-1}{{\bf Q}}^{{\rm III}} & \ \ \
\left[{{\bf Q}}^{{\rm IV}}\right]^{-1}
\end{array}
\right)\;. \label{eq:McRae2}
\end{eqnarray}
Multiplying both sides of Eq.~(\ref{eq:McRae1}) with the above matrix
we obtain Eq.~(\ref{eq:eig}).

\newpage
\begin{references}
\bibitem[\dag]{ntua} Also at the Department of Physics, National
Technical University of Athens, Zografou Campus, GR-157~73, Athens,
Greece.

\bibitem{Landau} L.~D.~Landau and E.~M.~Lifshitz,
{\em Theory of Elasticity,} 3rd Edition (Butterworth-Heinemann Ltd.,
Oxford, 1986).

\bibitem{Ashcroft} W.~Lamb, D.~M.~Wood and N.~W.~Ashcroft,
Phys. Rev. B {\bf 21}, 2248 (1980).

\bibitem{Joannop} J.~D.~Joannopoulos, R.~D.~Meade and J.~N.~Winn,
{\em Photonic Crystals} (Princeton University Press, Princeton, N.J.,
1995).

\bibitem{Soukoulis} {\em Photonic Band Gap
Materials}, edited by C.~M.~Soukoulis (Kluwer Academic, Dordrecht,
1996).

\bibitem{Yabl1} E.~Yablonovitch, Phys. Rev. Lett. {\bf 58}, 2059
(1987); E.~Yablonovitch, T.~J.~Gmitter, Phys. Rev. Lett. {\bf 63},
1950 (1989).

\bibitem{Yabl2} E.~Yablonovitch, J. Phys.: Condens. Matter {\bf 5},
2443 (1993); T.~F.~Krauss and R.~M.~De~La~Rue, Progr. Quant.
Electronics {\bf 23}, 51 (1999).

\bibitem{PT99} B.~G.~Levi, Phys. Today, {\bf 52}, No.~1, 17, (1999).

\bibitem{SJ91} S.~John, Phys. Today {\bf 44}, No.~5, 32 (1991).

\bibitem{PSheng} {\em Scattering and Localization of
Classical Waves in Random Media}, edited by P.~Sheng (World
Scientific, Singapore, 1990).

\bibitem{Sig2} E.~N.~Economou and M.~Sigalas, J. Acoust. Soc. Am.
{\bf 95}, 1734 (1994); M.~Kafesaki, M.~M.~Sigalas, and E.~N.~
Economou, Solid State Commun. {\bf 96}, 285 (1995).

\bibitem{Kushw1} M.~S.~Kushwaha and B.~Djafari-Rouhani, J. Appl.
Phys. {\bf 80}, 3191 (1996); M.~S.~Kushwaha and P.~Halevi, J. Acoust.
Soc. Am. {\bf 101}, 619 (1997).

\bibitem{Kushw2} M.~S.~Kushwaha, B.~Djafari-Rouhani, L.~Dobrzynski,
and J.~O.~Vasseur, Eur. Phys. J. B {\bf 3}, 155 (1998).

\bibitem{Sprik98} R.~Sprik and G.~H.~Wegdam, Solid State Commun.
{\bf 106}, 77 (1998).

\bibitem{Kushw3} M.~S.~Kushwaha, P.~Halevi, L.~Dobrzynski, and
B.~Djafari-Rouhani, Phys. Rev. Lett. {\bf 71}, 2022 (1993);
M.~S.~Kushwaha, P.~Halevi, G.~Mart\'{\i}nez, L.~Dobrzynski, and
B.~Djafari-Rouhani, Phys. Rev. B {\bf 49}, 2313 (1994); M.~M.~Sigalas
and E.~N. Economou, Phys. Rev. Lett. {\bf 75}, 3580 (1995);
M.~S.~Kushwaha, P.~Halevi, L.~Dobrzynski, and B.~Djafari-Rouhani,
Phys. Rev. Lett. {\bf 75}, 3581 (1995).

\bibitem{Kushw4} M.~S.~Kushwaha and P.~Halevi, Appl. Phys. Lett.
{\bf 64}, 1085 (1994); M.~S.~Kushwaha and P.~Halevi, Appl. Phys.
Lett. {\bf 69}, 31 (1996); M.~S.~Kushwaha, Appl. Phys. Lett. {\bf
70}, 3218 (1997).

\bibitem{Sig3} M.~Sigalas and E.~N.~Economou, Solid State Commun.
{\bf 86}, 141 (1993); M.~M.~Sigalas and E.~N.~Economou, J. Appl.
Phys. {\bf 75}, 2845 (1994); M.~M.~Sigalas and E.~N.~Economou,
Europhys. Lett. {\bf 36}, 241 (1996); M.~M.~Sigalas, J. Acoust. Soc.
Am. {\bf 101}, 1256 (1997); M.~M.~Sigalas, J. Appl. Phys. {\bf 84},
3026 (1998).

\bibitem{Vasseur} J.~O.~Vasseur, B.~Djafari-Rouhani, L.~Dobrzynski,
 M.~S.~Kushwaha, and P.~Halevi, J. Phys.: Condens. Matter {\bf
 6}, 8759 (1994); J.~O.~Vasseur, B.~Djafari-Rouhani, L.~Dobrzynski,
 and P.~A.~Deymier, J. Phys.: Condens. Matter {\bf 9}, 7327 (1997);
 J.~O.~Vasseur, P.~A.~Deymier, G.~Frantziskonis, G.~Hong,
 B.~Djafari-Rouhani, and L.~Dobrzynski, J. Phys.: Condens. Matter
 {\bf 10}, 6051 (1998).

\bibitem{Sanchez} R.~M\'artinez-Sala, J.~Sancho, J.~V.~S\'anchez,
V.~G\'omez, J.~Llinares, and F.~Meseguer, Nature (London) {\bf 378},
241 (1995); J.~V.~S\'anchez-P\'erez, D.~Caballero,
R.~M\'artinez-Sala, C.~Rubio, J.~S\'anchez-Dehesa, F.~Meseguer,
J.~Llinares, and F. G\'alvez, Phys. Rev. Lett. {\bf 80}, 5325 (1998);
F.~Meseguer, M.~Holgado, D.~Caballero, N.~Benaches,
J.~S\'anchez-Dehesa, C.~L\'opez, and J.~Llinares, Phys. Rev. B {\bf
59}, 12169 (1999).

\bibitem{Torres} F.~R.~Montero de Espinosa, E.~Jim\'enez, and
M.~Torres, Phys. Rev. Lett. {\bf 80}, 1208 (1998); M.~Torres,
F.~R.~Montero de Espinosa, D.~Garc\'{\i}a-Pablos, and N.~Garc\'{\i}a,
Phys. Rev. Lett. {\bf 82}, 3054 (1999).

\bibitem{Localiz} R.~L.~Weaver, Wave Motion {\bf 12}, 129 (1990);
R.~L.~Weaver, Phys. Rev. B {\bf 47}, 1077 (1993); L.~Ye, G.~Cody,
M.~Zhou, and P.~Sheng, Phys. Rev. Lett. {\bf 69}, 3080 (1992).

\bibitem{Biot} M.~A.~Biot, J. Appl. Phys. {\bf 33,} 1482 (1961);
K.~Attenborough, Phys. Rep. {\bf 82}, 179 (1982).

\bibitem{Schwartz} L.~M.~Schwartz and D.~L.~Johnson, Phys. Rev. B
{\bf 30}, 4302 (1984); D.~L.~Johnson, T.~J.~Ploma and H.~Kojima, J.
Appl. Phys. {\bf 76}, 115 (1994).

\bibitem{Marqusee} J.~A.~Marqusee and J.~M.~Deutch, J. Chem. Phys.
{\bf 75}, 5239 (1981).

\bibitem{Johnson} D.~L.~Johnson, J. Chem. Phys. {\bf 77}, 1531 (1983).

\bibitem{Colloid} J.~Liu, L.~Ye, D.~A.~Weitz, and P.~Sheng, Phys. Rev.
Lett. {\bf 65}, 2602 (1990); X.~Jing, P.~Sheng, and M.~Zhou, Phys.
Rev. Lett. {\bf 66}, 1240 (1991); X.~Jing, P.~Sheng, and M.~Zhou,
Phys. Rev. A {\bf 46}, 6513 (1992).

\bibitem{Pendry92} J.~B.~Pendry and A.~MacKinnon, Phys. Rev. Lett.
{\bf 69}, 2772 (1992); J.~B.~Pendry, J. Mod. Opt. {\bf 41}, 209
(1994).

\bibitem{SKM92} N.~Stefanou, V.~Karathanos, and A.~Modinos, J. Phys.:
Condensed Matter {\bf 4}, 7389 (1992);  N.~Stefanou, V.~Yannopapas,
and A.~Modinos, Comput. Phys. Commun. {\bf 113}, 49 (1998).

\bibitem{Ohtaka} K.~Ohtaka, Phys. Rev. B {\bf 19}, 5057 (1979);
K.~Ohtaka and Y.~Tanabe, J. Phys. Soc. Jpn. {\bf 65}, 2276 (1996).

\bibitem{Leung2} Y.~Qiu, K.~M.~Leung, L.~Carin, and D.~Kralj, J.
Appl. Phys. {\bf 77}, 3631 (1995).

\bibitem{footnote1} The reader who is not familiar with on-shell methods may find it
easier to follow the formalism of the present paper if he
familiarizes himself with the concepts and methodology of these
methods as applied to the simpler problem of electron scattering in
crystals (see, e.g., chapters 3 and 4 of Ref.~\onlinecite{Modinos}).

\bibitem{Brill} D.~Brill and G.~Gaunaurd, J. Acoust. Soc. Am.
{\bf 81}, 1 (1987).

\bibitem{Chew} W.~C.~Chew, {\em Waves and Fields in
Inhomogeneous Media} (Van Nostrant Reinhold, New York, 1990).

\bibitem{Pendry} J.~B.~Pendry, {\em Low Energy Electron
Diffraction} (Academic Press, London, 1974).

\bibitem{Modinos} A.~Modinos, {\em Field, Thermionic and Secondary
Electron Emission Spectroscopy} (Plenum Press, New York, 1984).

\bibitem{Auld} B.~A.~Auld, {\em Acoustic Fields and Waves in Solids,}
Volume I (John Wiley \& Sons Inc., New York, 1973).

\bibitem{K98} V.~Karathanos, J. Mod. Opt. {\bf 45}, 1751 (1998).

\bibitem{Mod87} A.~Modinos, Physica A {\bf 141}, 575 (1987).

\bibitem{Rose} M.~E.~Rose, {\em Elementary Theory of Angular Momentum}
(John Wiley \& Sons Inc., New York, 1957).

\bibitem{McRae} E.~G.~McRae, Surf. Sci. {\bf 11}, 479 (1968); Surf. Sci. {\bf 11}, 492
(1968).
\end{references}
\end{document}